\documentclass[10pt,aps,prd,amsmath,amssymb,nofootinbib,superscriptaddress,preprintnumbers,hidelinks,eprint,
nobibnotes,
floatfix
]{revtex4-2}

\usepackage{ragged2e}
\usepackage{etoolbox}
\usepackage{orcidlink}
\usepackage{tabularray}
\usepackage{graphicx}% Include figure files
\usepackage{dcolumn}% Align table columns on decimal point
\usepackage{bm}% bold math
\usepackage{cancel}
\usepackage{array}
\usepackage{array,url,kantlipsum}
\usepackage{amsmath}
\usepackage{amssymb}

\usepackage{mathabx}
\usepackage{mathrsfs}  
\usepackage{tensor}
\usepackage{comment}
\usepackage[normalem]{ulem}
\usepackage{cancel}
\usepackage{booktabs}
\usepackage{caption}
\usepackage[caption=false]{subfig}
\usepackage{subcaption}

\usepackage{graphicx}
\usepackage{dcolumn}
\usepackage{xcolor,colortbl}
\usepackage{multirow}
\usepackage{enumitem}

\usepackage{multirow}

\usepackage{hyperref}
\usepackage{stackengine,scalerel}
\hypersetup{colorlinks, linkcolor={red},citecolor={blue},urlcolor={blue}}  

\usepackage{physics,mathtools,tensor}
\usepackage{soul}
\usepackage{verbatim}

\newcommand{\rs}{r_0}

\newcommand{\omell}{\omega_\ell}

\renewenvironment{widetext@grid}{%
  \par\ignorespaces
  \setbox\widetext@top\vbox{%
   \vskip15\p@
   \hb@xt@\hsize{%
    \leaders\hrule\hfil
    \vrule\@height6\p@
   }%
   \vskip6\p@
  }%
  \setbox\widetext@bot\hb@xt@\hsize{%
    \vrule\@depth6\p@
    \leaders\hrule\hfil
  }%
  \onecolumngrid
  \let\set@footnotewidth\set@footnotewidth@ii
}{%
  \par
  \twocolumngrid\global\@ignoretrue
  \@endpetrue
}%
\makeatother

\makeatletter
\long\def\@makecaption#1#2{%
  \par
  \vskip\abovecaptionskip
  \begingroup
    \small\rmfamily
    \samepage
    \flushing
    \let\footnote\@footnotemark@gobble
    \@make@capt@title{#1}{#2}\par
  \endgroup
  \vskip\belowcaptionskip
}
\makeatother

\begin{document}

\newcommand{\ellp}{\ell_\mathrm{P}}
\newcommand{\gnewton}{G_\mathrm{N}}

\def\checkmark{\tikz\fill[scale=0.4](0,.35) -- (.25,0) -- (1,.7) -- (.25,.15) -- cycle;}

\numberwithin{equation}{section}
\renewcommand{\theequation}{\arabic{section}.\arabic{equation}}

\title{\bf \Large 
Analytic backreaction of a scalar wig on a Schwarzschild black hole}

\author{Marco de Cesare {\Large \orcidlink{0000-0002-4263-1009}}\,}
\email[]{marco.decesare@na.infn.it}
\affiliation{Scuola Superiore Meridionale, Largo S. Marcellino, 10, 80138 Napoli, Italy.}
\affiliation{INFN sezione di Napoli, via Cintia, 80126 Napoli, Italy.}

\author{Manuel Del Piano {\Large \orcidlink{0000-0003-4515-8787}}\,}
\email[]{madelp@qtc.sdu.dk}
\affiliation{Quantum  Theory Center ($\hbar$QTC) \& D-IAS, Southern Denmark University, Campusvej 55, 5230 Odense M, Denmark.}

\author{Carlos A. R. Herdeiro {\Large\orcidlink{0000-0002-9619-2013}}\,}
\email[]{herdeiro@ua.pt}
\affiliation{Departamento de Matemática da Universidade de Aveiro and Center for Research and Development
in Mathematics and Applications (CIDMA), Campus de Santiago, 3810-193 Aveiro, Portugal.}
\affiliation{Programa de Pós-Graduação em Física, Universidade Federal do Pará, 66075-110, Belém, Pará, Brazil.}

\begin{abstract}
We analytically determine the leading backreaction of a spherically symmetric massive complex scalar quasi-bound state (with mass $\mu$) on a Schwarzschild black hole with (initial) gravitational radius $r_0$. Working in the small-coupling regime, $\rs \mu \ll 1$, we evaluate the stress-energy tensor of the fundamental scalar $s$-wave and solve the Einstein equations through quadratic order in its amplitude in ingoing Eddington–Finkelstein coordinates. 
We also determine the small-mass quasi-resonant frequency of the fundamental $s$-wave analytically by matched asymptotic expansions and validate it numerically using Leaver's method.
Unlike steady-state treatments, the calculation retains the exponential decay of the quasi-bound state. We obtain explicit expressions for the metric perturbations and Misner–Sharp mass and derive the evolution of the future outer trapping horizon. The black-hole mass grows monotonically with the decaying horizon flux and saturates when the finite scalar cloud has been absorbed, with the decrease of the cloud mass exactly balancing the horizon growth at the perturbative order considered. We also determine the domain in which the scalar small-coupling approximation and the gravitational perturbative expansion are simultaneously valid.
\end{abstract}

\maketitle

\nopagebreak

\section{Introduction}
Light bosonic fields represent a promising cold dark matter candidate. In particular, ultralight (pseudo)scalar fields with sub-eV masses, 
corresponding to a range of Compton wavelengths interesting for various astrophysical phenomena, give rise to interesting phenomenology and have been extensively investigated within the context of the fuzzy dark matter scenario \cite{Hu:2000ke,Hui:2016ltb,Hui:2021tkt,Eberhardt:2025caq}.
Such scalar fields may form non-trivial environments around black holes~\cite{Barranco:2012qs,Herdeiro:2014goa} and other compact objects, as well as solitonic cores in galactic dark matter haloes \cite{Bar:2018acw}, and stable self-gravitating configurations such as boson stars \cite{Liebling:2012fv} or Proca stars~\cite{Brito:2015pxa}.

The existence of \textit{strictly} equilibrium scalar field configurations around asymptotically flat black holes in general relativity is restricted by several ``no-hair" theorems~\cite{Herdeiro:2015waa}. However, these obstructions are evaded if the scalar field is time dependent.
The existence of long-lived dynamical scalar field solutions supported by a Schwarzschild black hole, known as `scalar wigs', has been established in Refs.~\cite{Barranco:2011eyw,Barranco:2012qs}. Such configurations are also known in the literature as `quasi-resonances' \cite{Ohashi:2004wr} or `gravitational atoms', and can be generalized to a Kerr background \cite{Baumann:2019eav}. 
Although these solutions are obtained in the test-field approximation, they are consistent with the long-term evolution of the full nonlinear Einstein-Klein-Gordon system found in numerical studies ~\cite{Okawa:2014nda,Sanchis-Gual:2014ewa,Barranco:2017aes}. Scalar wigs around black holes may form via different mechanisms, such as superradiance (in the rotating case) \cite{Brito:2015oca} and accretion from the galactic environment \cite{Hui:2019aqm,Budker:2023sex,Cardoso:2022nzc}, also in the static case.

In the stationary case, strict equilibrium turns out to be possible. There exists a family of `hairy' rotating black hole solutions with a complex scalar field, where the scalar has a harmonic time dependence and is subject to a suitable synchronization condition with the horizon angular velocity \cite{Herdeiro:2014goa}. 
The accretion of a ultralight scalar field onto a black hole with synchronized scalar hair has been studied numerically using the Bondi-Hoyle-Lyttleton model in Ref.~\cite{Cruz-Osorio:2023wev}.
Other examples of non-trivial time-dependent scalar field configurations around black holes have been studied in Refs.~\cite{Clough:2019jpm}. 
Such non-trivial matter environments around black holes leave distinctive imprints on gravitational waves, which are particularly relevant for signals emitted from extreme mass ratio inspirals \cite{Degollado:2014vsa,Collodel:2021jwi,DellaRocca:2024sda,Duque:2023seg,Brito:2023pyl,Mariano:2025gyp}, induce frequency shifts in the quasinormal mode spectrum \cite{DellaRocca:2025xwz}, and also have dynamical effects on the black hole shadow \cite{Acevedo-Munoz:2025ueh}. Moreover, the accretion rate of environmental scalar fields is sensitive to deviations from the standard purely ingoing boundary conditions at the black-hole horizon, which may arise due to quantum gravity effects \cite{Mitra:2023sny}.

In this work we will focus on a Schwarzschild black hole endowed with a spherically symmetric `scalar wig'. As the scalar field decays in time, it is absorbed by the black hole and backreacts on the geometry. This process has been previously studied numerically or in the test field approximation~\cite{Barranco:2012qs,Barranco:2013rua,Barranco:2017aes,Aguilar-Nieto:2022jio}; here we perform an \textit{analytical} calculation of the backreaction, solving the Einstein field equations perturbatively in Eddington-Finkelstein coordinates. In particular, we compute the time-dependent Misner-Sharp mass of the perturbed geometry and the evolution of the black-hole apparent horizon. 

Our approach to compute the backreaction in spherical symmetry generalizes Ref.~\cite{Babichev:2012sg},\footnote{The approach of Ref.~\cite{Babichev:2012sg} has been further extended to slowly rotating black holes and matter fields with a non-trivial angular dependence
%(though still in a steady-state approximation)
in Ref.~\cite{Kimura:2021dsa}; however, the application to the Blandford–Znajek considered there still relies on a steady-state approximation. } where the backreaction of matter (specifically, a perfect fluid and a massless scalar field) onto a black hole  is computed under the simplifying assumption of a steady-state accretion. However, in the case of a scalar wig profile the steady-state approximation does not hold, due to the fact that the scalar field undergoes an exponential time decay. Therefore, a more general treatment is required where full time dependence is retained, which will be presented in this work. The backreaction of a scalar cloud has been computed in Refs.~\cite{Bamber:2021knr,DeLuca:2021ite} treating the cloud as stationary, which results in a steady-state accretion process. Although the cloud may be well approximated as stationary over short timescales compared to the cloud lifetime, this approximation is unable to capture the entire accretion process and the relaxation of the system to the final state where the cloud has been completely absorbed by the black hole. Our approach, on the other hand, enables this.

The paper is organized as follows. In Section~\ref{Sec:2} we review massive scalar wig solutions on a Schwarzschild background, focusing on the spherically symmetric case, and compute the corresponding stress-energy tensor. The scalar wig is treated analytically in the small-field limit.
In Section~\ref{sec-mass-function} we solve the Einstein field equations perturbatively in Eddington-Finkelstein coordinates, computing the metric perturbations explicitly. We also compute analytically the evolution of the apparent horizon and the Misner-Sharp mass of the perturbed geometry. Two technical appendices are included. 
Appendix~\ref{AppDetweiler} provides further mathematical details on the low-mass approximation. We first review the Detweiler approximation, whose extrapolation to $s$-wave modes in Schwarzschild has been used in several works, and then improve on it by computing the quasi-resonant frequency spectrum with two different methods: analytically using matched asymptotic expansions, and numerically using Leaver's method.
In Appendix~\ref{Appintegrals} we report the detailed analytical computation of the integrals appearing in the solution for the Misner-Sharp mass.

\section{Dynamical resonances of a massive complex scalar field}\label{Sec:2}

Let us consider a complex scalar field $\Phi$ of mass $\mu$ on a Schwarzschild background. 
We consider the following line element for a spherically symmetric geometry in Schwarzschild-type coordinates
\begin{equation}\label{metric-t-r}
    \mathrm{d}s^{2}
    = -\left(1-\frac{2GM(t,r)}{r}\right)\mathrm{d}t^{2}\\
    + \left(1-\frac{2GM(t,r)}{r}\right)^{-1}\mathrm{d}r^{2}
    + r^{2}\mathrm{d}\Omega^{2}\,,
\end{equation}
where $r$ is the areal radius, $\dd \Omega^2$ is the line element of $S^2$ and $G$ is Newton's gravitational constant which is related to Planck's mass as $G=M_{\rm P}^{-2}$. We assume a dynamical spacetime through the time dependence of the mass function $M(t,r)$, which includes the (initial) background black-hole mass $M_{0}$ and small backreaction corrections due to the scalar field accretion, so that $|M(t,r)-M_{0}|\ll M_{0}$.

Let us consider a massive complex scalar field minimally coupled to gravity, described by the action
\begin{equation}\label{Lagr}
    S=\int \dd^4 x \sqrt{-g} \left[ \frac{R}{16 \pi G } - \nabla_\rho \overline{\Phi} \nabla^\rho \Phi - \mu^2 |\Phi|^2 \right] \ ,
\end{equation}
where the overline indicates complex conjugation. The dynamics are given by the Einstein and Klein-Gordon field equations
\begin{equation}\label{KG equation}
    R_{\mu \nu}-\frac{R}{2}g_{\mu \nu} = 8 \pi G\, T_{\mu \nu} \qq{and} (\nabla^\nu \nabla_\nu - \mu^2 )\Phi = 0 \ ,
\end{equation}
where the energy-momentum tensor for the scalar is
\begin{equation}\label{EMT def}
    T_{\mu \nu} = \nabla_\mu \overline{\Phi}  \nabla_\nu \Phi + \nabla_\nu \overline{\Phi}  \nabla_\mu \Phi - g_{\mu \nu} \left( \nabla_\rho \overline{\Phi} \nabla^\rho \Phi + \mu^2 |\Phi|^2\right) ~.
\end{equation}

To leading order, the scalar field can be treated in the test-field approximation as propagating on a static Schwarzschild black hole with unperturbed mass $M_0$.
In the remaining of this section we focus on such unperturbed background. Then, we decompose the scalar field in spherical harmonics\footnote{The spherical symmetry allows to drop the dependence on the subscript $m$ in $\omega$ and $\Phi_\ell$.}
\begin{equation}\label{scalar field def}
    \Phi(t,r,\theta,\phi)= \sum_{{-\ell \leq m \leq \ell}\atop{\ell \geq 0}} \frac{\Phi_{\ell}(t,r)}{r}Y_{\ell}^m(\theta,\phi) \ .
\end{equation}
The partial waves $\Phi_{\ell}$ depend on both $t$ and $r$. Next, we transform to the frequency domain, using the following conventions for the Fourier transform, $\Phi_{\ell}(t,r)=\frac{1}{\sqrt{2\pi}}\int_{-\infty}^{+\infty}\dd \omell\, e^{- i \omell t}  \Psi_{\ell}(\omega_{\ell},r)$~. The resulting equations for the various monochromatic components read as
\begin{equation}\label{regge-wheeler-eq}
    \pdv[2]{\Psi_\ell}{r_\ast} + \big( \omell^2 - V_\ell(r) \big)\Psi_\ell = 0 \ ,
\end{equation}
where we have introduced the \textit{tortoise coordinate} and the effective potential, which are defined, respectively, as
\begin{subequations}
\begin{align}
    &r_\ast \coloneqq r + \rs \ln\left(\frac{r}{\rs} - 1 \right) \ , \\ 
    &V_\ell(r) \coloneqq f(r)\left(\frac{\ell(\ell+1)}{r^2} + \frac{\rs}{r^3} + \mu^2 \right) \ .
\end{align}
\end{subequations}
Here, for notational convenience, we introduced $f(r)\coloneqq 1-\rs/r$ and $\rs \coloneqq 2GM_0$, corresponding to the Schwarzschild lapse function and the event horizon radius of the background black hole, respectively. 
We make the following choice of boundary conditions, following Ref.~\cite{Barranco:2012qs},
\begin{equation}\label{Eq:PsiAsymptotics}
\Psi_\ell \sim
    \begin{cases}
        e^{- i \omell r_\ast} & \qq{for} r_\ast \to - \infty ~, \\
        e^{- \chi_\ell r_\ast} r^{ r_0 \mu^2 / 2\chi_\ell} & \qq{for} r_\ast \to + \infty ~,
    \end{cases}
\end{equation}
with $\chi_\ell \coloneqq \sqrt{\mu^2 - \omell^2}$~. Note that $\chi_\ell$ has a positive real part, which ensures exponential decay at spatial infinity.
Moreover, we assume a single monochromatic component with frequency $\omega_\ell$ in the quasi-resonant spectrum~\cite{Barranco:2012qs,Konoplya:2004wg}.
Hence, we make the following ansatz in real space (cf.~Refs.~\cite{Konoplya:2004wg,Konoplya:2011qq,Leaver:1985ax,Nollert1993})
\begin{equation}\label{psi def}
    \Psi_\ell(r) = e^{- \chi_\ell r} \left(\frac{r}{\rs}\right)^{-\rs [\chi_\ell - \mu^2/(2\chi_\ell)]} \left(1 - \frac{\rs}{r} \right)^{- i \rs \omell} S(r) \ ,
\end{equation}
where we define
\begin{equation}\label{S definition}
    S(r) \coloneqq \sum_{k\geq 0} a_k \left(1 - \frac{\rs}{r} \right)^k \ ,
\end{equation}
in which the infinite sum extends over non-negative integers $k$\,. Inserting the expression \eqref{psi def} into Eq.~\eqref{regge-wheeler-eq}, we obtain a three-term recursive relation\footnote{Note that our definition of $\chi_\ell$ follows Ref.~\cite{Barranco:2012qs}, which is different from the one used in Ref.~\cite{Konoplya:2004wg}.}
\begin{align}\label{leaver-recursive-relation}
    &\alpha_0 a_1 + \beta_0 a_0 = 0 \ , \notag \\ 
    & \alpha_{n} a_{n+1} + \beta_{n} a_n + \gamma_{n} a_{n-1} = 0  \qq{with} n \geq 1 \ ,
\end{align}
where $a_0$ is a free parameter controlling the amplitude of the solution in the linear regime. One may factor this term out and identify it with a small book-keeping parameter $\epsilon$, corresponding to the overall amplitude of the test field. In practice, to avoid overburdening the notation, one sets $a_0=1$ in the expressions while keeping track of the perturbative order. The overall amplitude of the test field may be easily reinstated at a later stage.
The solution to the recursive relations \eqref{leaver-recursive-relation} reads \cite{Konoplya:2004wg,Konoplya:2011qq}
\begin{subequations}\label{leaver-coefficients}
\begin{align}
    \alpha_n &= (n+1)(n + 1 - 2 i \rs \omega_\ell) \ ,\\
    \beta_n &= \frac{\rs(\omega_\ell + i \,\chi_\ell) (2\rs (\omega_\ell + i \, \chi_\ell)^2 + i (2 n +1)(\omega + 3 i \, \chi_\ell)}{2i \, \chi_\ell} - 2n(n+1) - 1 - \ell (\ell+1) \ , \\
    \gamma_n & = \left(n - \, \rs \frac{(\omega_\ell + i \, \chi_\ell)^2}{2 \, \chi_\ell} \right)^2 \ .
\end{align}
\end{subequations}
To determine the eigenvalues $\omega_\ell$ using Leaver's method, one has to determine the roots of the continued fraction that derives from the ratio $a_{n+1}/a_n$ of the coefficients of the series $S(r)$, which is written in terms of the recursive relation coefficients in Eq.~\eqref{leaver-recursive-relation}.

To ensure that perturbations preserve the spherical symmetry of the background, let us now restrict our analysis to the fundamental $s$-wave mode $(n,\ell)=(0,0)$ and drop indices from $\omega$ and $\chi$\,. Using Eqs.~\eqref{Eq:spectrum_mae}, and within a low-mass approximation $GM_0 \mu  \sim \rs \mu \ll 1$, the frequency of the $s$-wave is given by 
\begin{subequations}\label{detweilers approx_zerospin}
\begin{align}\label{detweiler-re-im-omega-swave}
\Re(\omega) &= \mu \sqrt{ 1 - \left(\frac{\rs\mu}{2}\right)^2} \qq{and} \Im( \rs \, \omega) = - \frac{( \rs \mu)^6}{4} \ ,
\end{align}
\end{subequations}
so that, assuming $\rs \mu \ll 1$, we have
\begin{subequations}
    \begin{align}\label{detweiler-omega2-chi}
    |\omega|^2 &= \mu^2\left[1-(\rs \mu)^2/4 +\order{\rs \mu}^{10} \right] \ , \\
    \chi &= \mu \left[\rs \mu/2 + i (\rs \mu)^4/2 -  i (\rs \mu)^6/16 + (\rs \mu)^7/4 + \order{\rs \mu}^8 \right] \ .  \label{detweiler-chi}
    \end{align}
\end{subequations}
At this point, one can use the above results and the recursive relations~\eqref{leaver-recursive-relation} to compute the coefficients in the power series $S(r)$ in Eq.~\eqref{S definition}, and here we report the leading orders in powers of $\rs \mu$ of the coefficients $a_i$ for $i=1,2,3$
\begin{subequations}
\begin{align}\label{detweiler-a_n}
    a_1 &=  - 2 ( \rs \mu)^2 - 5 i \, (\rs \mu)^3 + 13 (\rs \mu)^4 + \frac{221}{8} i \, (\rs \mu)^5 +\order{\rs \mu}^6 \ ,\\
    a_2 &= - \frac{3}{4} (\rs \mu)^2 -  i \, (\rs \mu)^3 + \frac{15}{8} (\rs \mu)^4 + \frac{7}{4} i \, (\rs \mu)^5 + \order{\rs \mu}^6\ \, ,\\
    a_3 &= -\frac{4}{9} (\rs \mu)^2 - \frac{11}{27}i \, (\rs \mu)^3 + \frac{89}{162}(\rs \mu)^4  - \frac{323}{1944}i \, (\rs \mu)^5 + \order{\rs \mu}^6 \ .
\end{align}
\end{subequations}
Interestingly, the real part of the coefficients is determined by the even powers of $\rs \mu$\,, whereas the imaginary part arises from the odd powers.

Going back to the original $\Phi$ variable, and restricting Eq.~\eqref{scalar field def} to the fundamental $s$-wave mode, we obtain
\begin{equation}
    \Phi(t,r)= \epsilon \, e^{- i \omega t}\frac{ \Psi(r) }{\sqrt{8 \pi^2}\,  r}  \ ,
\end{equation}
with $\Psi$ given by Eq.~\eqref{psi def} and $\epsilon$ is the dimensionless parameter controlling the amplitude of $\Phi$.
Now, we transform to (ingoing) Eddington-Finkelstein coordinates $(v,r)$\,, with $v \coloneqq t + r_\ast$\,, which are regular at the horizon. 
In these coordinates, the background metric reads
\begin{equation}\label{EF metric}
    \dd s^2 = - \left( 1 - \frac{\rs}{r}\right) \dd v^2 + 2  \dd v \, \dd r + r^2 \dd \Omega^2 \ .
\end{equation}
Since $\Phi$ is a scalar, we simply obtain
\begin{equation}\label{phi spherical}
    \Phi(v,r)= \epsilon \, e^{- i \omega (v-r_\ast)} \frac{ \Psi(r)}{\sqrt{8 \pi^2} \, r} \ .
\end{equation}

\begin{figure*}[t]
  \centering
  \includegraphics[width=\textwidth]{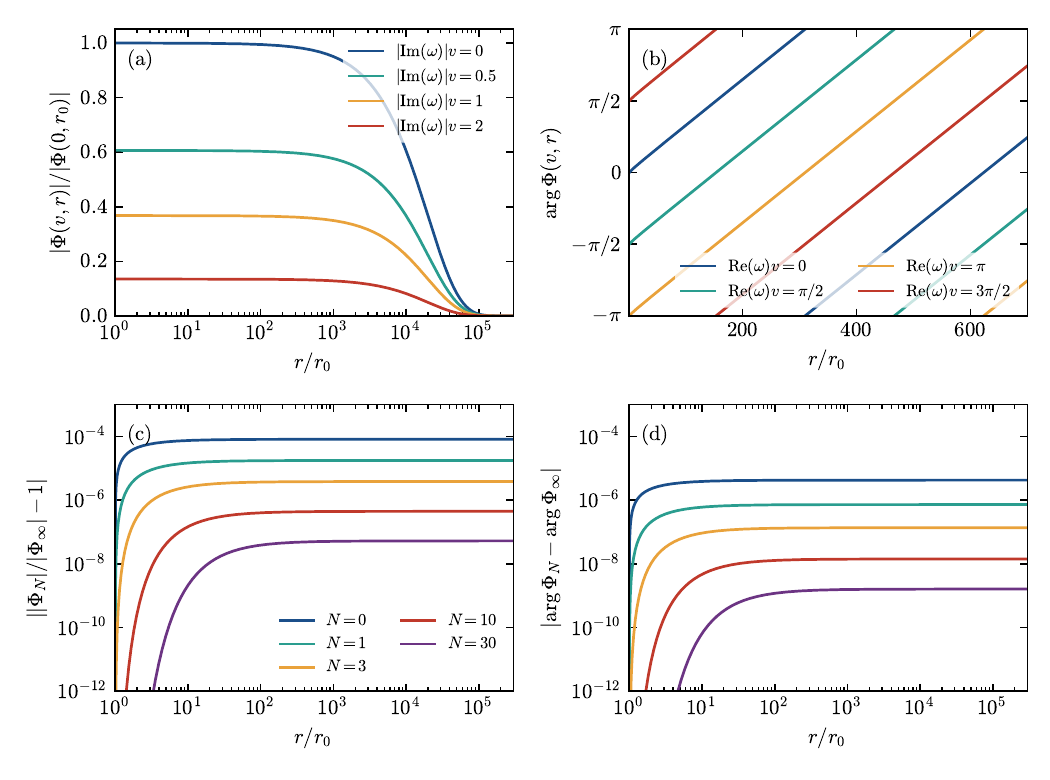}
  \caption{Fundamental ($n=0$, $\ell=0$) quasi-bound state of the test scalar field for $r_0\mu=10^{-2}$, a value lying safely within the small-coupling regime assumed in our analytical treatment in Sec~\ref{sec-mass-function}, in ingoing Eddington--Finkelstein coordinates, Eq.~\eqref{phi spherical}. The frequency, $r_0\omega = 9.99987\times10^{-3} - 2.5088\times10^{-13}\, i$, is obtained numerically from Leaver's continued fraction. The global $U(1)$ phase is fixed by requiring $\Phi(0,r_0)$ to be real and positive, and the field is normalized to its horizon value at $v=0$. (a)~Modulus at different advanced times, in units of the decay time $|\Im(\omega)|^{-1}\approx 4.0\times10^{12}\,r_0$: the radial shape is preserved and the amplitude decays as $e^{\Im(\omega) v}$. (b)~Phase at four instants within one oscillation period $2\pi/\Re(\omega)\approx 628\,r_0$: the profile is rigidly shifted by $-\Re(\omega) v$. (c),(d)~Relative deviation in modulus and phase of the profile computed with the truncated series $S_N(r)$ from the converged one, $\Phi_\infty$ (computed with $10^{6}$ terms). The deviations vanish at the horizon and saturate for $r\gtrsim 10\,r_0$.}
  \label{fig:wig_profile}
\end{figure*}

The scalar wig in the test-field approximation is shown in Fig.~\ref{fig:wig_profile} for $r_0\mu=10^{-2}$, using the frequency computed numerically with Leaver's method, which agrees with the small-mass expression~\eqref{detweiler-re-im-omega-swave} to $3\times10^{-9}$ in $\Re(\omega)$ and $0.4\%$ in $\Im(\omega)$. The dependence on advanced time at fixed $r$ factorizes as
\begin{equation}
  |\Phi(v,r)| = e^{\Im(\omega) v}\,|\Phi(0,r)|\ , \qquad
  \arg\Phi(v,r) = \arg\Phi(0,r) - \Re(\omega) v\ ,
\end{equation}
so that, at this order, the radial shape of the modulus is preserved while its amplitude decays on the timescale $|\Im(\omega)|^{-1}$, which exceeds the oscillation period $2\pi/\Re(\omega)$ by about ten orders of magnitude for $r_0\mu=10^{-2}$. The modulus is nearly constant near the horizon and decays exponentially on the scale set by $\chi^{-1}\approx 2/(r_0\mu^2)$, reaching half of its horizon value at $r\approx 1.4\times10^{4}\,r_0$. On slices of constant $v$ the phase grows approximately as $\Re(\omega) r_\ast$ at large $r$; equivalently, $\arg\Phi\approx-\Re(\omega) t$ up to small corrections, reflecting the standing-wave character of the quasi-bound state in Schwarzschild time. Non-trivial distortions of the radial profile due to the evolving geometry arise only beyond the perturbative order considered here.
 
The lower panels of Fig.~\ref{fig:wig_profile} show the effect of truncating the series $S(r)$ at finite order,
\begin{equation}
  S_N(r)=\sum_{k=0}^{N} a_k\left(1-\frac{r_0}{r}\right)^{k}.
\end{equation}
Since the expansion variable $1-r_0/r$ vanishes at the horizon, the terms with $k>0$ do not modify the horizon value of the field, and their effect grows away from the horizon, approaching its asymptotic value at increasingly large radii as $N$ increases. This behavior is consistent with the leading neglected term being proportional to $(1-r_0/r)^{N+1}$, which approaches its asymptotic value for $r\gg (N+1)r_0$. The series converges rapidly: for $r_0\mu=10^{-2}$ the leading term alone reproduces the modulus to within $10^{-4}$ and the phase to within $5\times10^{-6}$\,rad at all radii, and each further order reduces the deviation, $N=10$ bringing it below $5\times10^{-7}$. The overall shape of the wig is therefore already captured by the prefactor multiplying $S(r)$ in Eq.~(2.18), the higher-order terms providing small corrections.

Using Eq.~\eqref{phi spherical}, we obtain the following expressions for the components of the gradient of the scalar field
\begin{equation}
    \partial_v \Phi = - i \omega \Phi \qq{and}  \partial_r \Phi = Z(r) \Phi  \ ,
\end{equation}
where, as a convenient short-hand notation, we introduced the following auxiliary function
\begin{equation}\label{z-definition}
    Z(r) \coloneqq \zeta  + \xi \, \frac{\rs}{r}   + \dv{\log S(r)}{r} \ ,
\end{equation}
where we have defined the constants
\begin{equation}\label{zeta-xi-def}
    \zeta \coloneqq i \omega - \chi \qq{and} \xi \coloneqq \zeta + \frac{\mu^2}{2 \chi} - \rs^{-1} \ .
\end{equation}
The non-vanishing components of the energy-momentum tensor \eqref{EMT def} read
\begin{subequations}\label{Tmunu}
    \begin{align}
        \tensor{T}{_v^v}& = -\left( \mu^2 + f(r)|Z(r)|^2 \right) |\Phi|^2 \ , \label{Tvv}\\
        \tensor{T}{_r^v}&=  2|Z(r)|^2  |\Phi|^2 \ , \label{Tvr}\\
        \tensor{T}{_v^r}&=  2\left(|\omega|^2 + f(r) \Im \left( \overline{\omega} Z(r)\right)\,\right)|\Phi|^2 \ , \label{Trv}\\
        \tensor{T}{_r^r}&= \left( f(r) |Z(r)|^2 - \mu^2\right)|\Phi|^2 \ ,\\
    \tensor{T}{_\theta^\theta}&=\tensor{T}{_\phi^\phi} =  \left[ 2\Im \left( \overline{\omega} Z(r)\right) - f(r) |Z(r)|^2 - \mu^2 \right] |\Phi|^2 ~ .
    \end{align}
\end{subequations} 
We remark that all of the above components are functions of the advanced time $v$ through $|\Phi(v,r)|$ ,
\begin{equation}\label{phi-modulus-squared}
     |\Phi(v,r)|^2 = \epsilon^2 \frac{e^{2 \Im(\omega) v}}{8 \pi^2}  \frac{e^{- \mathcal{A} \, r/ \rs}}{r^2} \left( \frac{r}{\rs}\right)^{\mathcal{B}}|S(r)|^2 \ ,
\end{equation}
where we have introduced
\begin{subequations}
\begin{align}
    \mathcal{A} & \coloneqq 2 \rs \Re\left( \chi  - i \, \omega  \right) \ , \label{curlyAdef}\\
    \mathcal{B} & \coloneqq  2 \rs \Re\left( i \, \omega - \chi + \frac{ \mu^2}{2\chi}\right) \label{curlyBdef}\ .
\end{align}
\end{subequations}
The time-dependence of the energy-momentum tensor stems from the fact that ${\rm Im}(\omega)\neq0$\,.

\section{Backreaction and black-hole accretion law}\label{sec-mass-function}

To compute the gravitational backreaction of the scalar wig, it is convenient
to work in ingoing Eddington--Finkelstein coordinates, which are regular at the
future event horizon. Accordingly, we adopt the following spherically symmetric
metric ansatz,
\begin{equation}
\dd s ^2=-e^{2\lambda(v,r)}
\left(1-\frac{2GM(v,r)}{r}\right)\dd v^2
+2e^{\lambda(v,r)}\dd v\,\dd r+r^2\dd\Omega^2 ,
\label{Babichev metric}
\end{equation}
where $r$ is the areal radius, while the functions $M(v,r)$ and
$\lambda(v,r)$ encode the backreaction of the scalar field on the geometry. {Specifically, $M(v,r)$ is the Misner-Sharp mass of the perturbed geometry.}
In the absence of matter, the metric reduces to the Schwarzschild solution,
with
\begin{equation}
M(v,r)=M_0~,
\qquad
\lambda(v,r)=0~.
\end{equation}

The metric above still possesses a residual gauge freedom associated with
reparametrizations of the null coordinate,
$v\rightarrow\tilde v(v)$.
Under such transformations, the function $\lambda$ is shifted by an arbitrary
function of $v$, which can be fixed by an appropriate gauge choice. In these coordinates, the Einstein field equations reduce to~\cite{Babichev:2012sg}\footnote{However, note the different signature convention compared to Ref.~\cite{Babichev:2012sg}.}
\begin{subequations}\label{Eq:efe}
\begin{align}
    \partial_r M & = - 4 \pi r^2 \, \tensor{T}{_v^v} = 4 \pi r^2 \rho(v,r)\ , \label{BabichevMprime}\\ 
    \partial_v M & =  4 \pi r^2 \, \tensor{T}{_v^r} =  \mathcal{F}(v,r)\ , \label{BabichevMdot}\\
    \partial_r\left(e^{-\lambda}\right) &=  4 \pi G\, r \, \tensor{T}{_r^v} \ ,\label{Babichevlambda}
\end{align}
\end{subequations}
where the energy density and the energy flux across a sphere of radius $r$ are defined, respectively, as follows
\begin{equation}\label{rEF-flux-definition}
    \rho(v,r)  \coloneqq - \tensor{T}{_v^v} > 0 \qq{and}  \mathcal{F}(v,r) \coloneqq 4 \pi r^2 \tensor{T}{_v^r} \ . 
\end{equation}
{Note that, since the scalar field is exponentially localized, the energy flux $\mathcal{F}$ tends to zero in the large-distance limit. Then, Eq.~\eqref{BabichevMdot} implies $\lim_{r \to \infty }\partial_v M = \lim_{r \to \infty }\mathcal{F}(v,r) = 0$\,. Therefore, the Misner-Sharp mass tends to a constant in this limit, which coincides with the total mass of the black-hole–scalar-field system.}
The Einstein equations~\eqref{Eq:efe} are exact for the metric ansatz~\eqref{Babichev metric}. In what follows, their source is evaluated using the test-field stress-energy tensor obtained in Sec.~\ref{Sec:2}, which is accurate through $\mathcal{O}(\epsilon^2)$.

In Ref.~\cite{Babichev:2012sg}, as well as in later works \cite{DeLuca:2021ite,Bamber:2021knr,Brito:2023pyl}, Eqs.~\eqref{BabichevMprime} and \eqref{BabichevMdot} are solved under a steady-state approximation, whereby the components of the energy--momentum tensor are taken to be independent of $v$. This assumption is not appropriate for the `scalar wig', which exhibits an explicit time dependence and decays exponentially on long timescales of order $T \sim 1/{\rm Im}(\omega)$\,. Since our aim is to compute the backreaction of the scalar wig on the geometry and to characterize the long-term evolution of the system, we shall drop the assumption of a steady-state evolution of matter. Therefore, in what follows, the functions $M$ and $\lambda$ are determined by solving the field equations taking into account the time dependence of $\tensor{T}{_\mu^\nu}$ (which has been previously computed in Eqs.~\eqref{Tmunu} in Section~\ref{Sec:2} in the test-field approximation).

We now proceed to solve the field equations~\eqref{Eq:efe}. First, we note that, differentiating Eq.~\eqref{BabichevMprime} with respect to $v$ and Eq.~\eqref{BabichevMdot} with respect to $r$ and using the Schwarz's theorem, we obtain the following condition, expressing the conservation of the stress-energy tensor
\begin{equation}
    0=-(4\pi)^{-1}\left(\partial_{v}\partial_{r} M-\partial_{r}\partial_{v} M\right)=r^2 \partial_v\, \tensor{T}{_v^v} + \partial_r (r^2 \, \tensor{T}{_v^r})  \ .\label{condition emt}
\end{equation}
More explicitly, the conservation of the stress-energy tensor in the coordinate system at hand can be expressed as
\begin{equation}
   \partial_r \mathcal{F} - 4 \pi r^2 \partial_v \rho  = 0 \ .
\end{equation}
Given the exponential dependence of $\rho$ in $v$\,, as in Eq.~\eqref{rEF-flux-definition},~\eqref{Tvv} and~\eqref{phi-modulus-squared}, we have
\begin{equation}
    \partial_r \mathcal{F}(v,r) = 8 \pi \Im(\omega) \,r^2 \rho \ .
\end{equation}
Integrating both sides in $r$, we have
\begin{equation}\label{Flux-1}
    \mathcal{F}(v,r) = 8 \pi \Im(\omega) \int_{r_0}^r \widetilde{r}^2 \rho(v,\widetilde{r}) \, \dd \widetilde{r} + \mathcal{F}(v,r_0)  \ .
\end{equation}
We also integrate Eq.~\eqref{BabichevMprime}, yielding
\begin{equation}\label{M-integral-r}
    M(v,r) = 4 \pi \int_{r_0}^r \widetilde{r}^2 \rho(v,\widetilde{r}) \, \dd \widetilde{r}  +  h(v) \ ,
\end{equation}
where $h(v)$ is an integration function. We can now insert Eqs.~\eqref{M-integral-r} and~\eqref{Flux-1} into Eq.~\eqref{BabichevMdot}, obtaining an equation for the integration function
\begin{equation}\label{diffeq-for-h}
    \partial_v h(v) = \mathcal{F}(v,r_0) \ .
\end{equation}
Combining Eqs.~\eqref{rEF-flux-definition} and~\eqref{Tvr}, we have that
\begin{equation}\label{flux-r0}
    \mathcal{F}(v,r_0)   = \epsilon^2\frac{ |\omega|^2 e^{-\mathcal{A}} }{\pi}e^{2 \Im(\omega) v} \ ,
\end{equation}
so Eq.~\eqref{diffeq-for-h} is integrated as
\begin{equation}\label{hv-sol}
    h(v) = \epsilon^2 \frac{|\omega|^2e^{2 \Im(\omega)v_0-\mathcal{A}}}{2 \pi  |\Im(\omega)|}\left(1-e^{2 \Im(\omega)(v-v_0)} \right) + h(v_0) \ ,
\end{equation}
where we used the fact that $\Im(\omega) <0$.
Hence, \eqref{M-integral-r} can be rewritten as
\begin{equation}\label{M-sol}
    M(v,r) = M_0 + 4 \pi \int_{r_0}^r \widetilde{r}^2 \rho(v,\widetilde{r}) \, \dd \widetilde{r} + \epsilon^2 \frac{|\omega|^2e^{2 \Im(\omega)v_0-\mathcal{A}}}{2 \pi |\Im(\omega)|}\left(1-e^{2 \Im(\omega)(v-v_0)} \right) \ ,
\end{equation}
where we fixed the integration constant $M(v_0,r_0) = h(v_0) = M_0$. The second term is the energy of matter at advanced-time $v$ in a spherical shell with outer radius $r$ and inner radius at the horizon $r_0$, whereas the third term represents matter accretion due to the influx of energy-momentum as the scalar wig decays in time.
Similar calculations for the backreaction have been performed in the literature before \cite{Brito:2023pyl,Bamber:2021knr,DeLuca:2021ite,Babichev:2012sg}, under more restrictive assumptions on the behaviour of matter. However, in Ref.~\cite{Brito:2023pyl} the integration functions that we report in \eqref{M-integral-r} and~\eqref{Flux-1} have been neglected, and in Refs.~\cite{Bamber:2021knr,DeLuca:2021ite,Babichev:2012sg} the accretion flux is assumed to be independent of $v$, which results in a linear $v$-dependence of the Misner-Sharp mass at finite $r$.
Lastly, Eq.~\eqref{Babichevlambda} is solved as 
\begin{equation}\label{lambda-sol}
    e^{-\lambda(v,r)} =  4 \pi G\int_{r_0}^{r} \tilde{r} \, \tensor{T}{_r^v}\dd\tilde{r} + K(v) = 8 \pi G \int_{\rs}^r  \widetilde{r}\,  |Z(\widetilde{r})|^2 |\Phi(v,\widetilde{r})|^2 \dd \widetilde{r}+ K(v) \ ,
\end{equation}
where we used the form of the component of the stress-energy tensor~\eqref{Tvr} and $K(v)$ is an arbitrary integration function arising from the radial
integration. Its presence reflects the residual gauge freedom associated
with reparametrizations of the null coordinate,
$v\rightarrow\tilde v(v)$, under which the metric retains the form~\eqref{Babichev metric} while the function $\lambda$ is shifted by an arbitrary function
of $v$. Accordingly, $K(v)$ can be fixed by a choice of normalization
for the null coordinate. In the following we adopt the convenient gauge
choice $K(v)=1$, 
which implies $e^{-\lambda(v,r_0)}=1$, 
so that the advanced time coordinate is normalized at the horizon.

We can express the solutions~\eqref{M-sol} and~\eqref{lambda-sol} as
\begin{equation}\label{M-sol-I}
    M(v,r)= M_0 + \frac{\epsilon^2}{2 \pi} \left[ e^{2\Im(\omega)v} \, \mathcal{I}(r) +\frac{|\omega|^2e^{2 \Im(\omega)v_0-\mathcal{A}}}{|\Im(\omega)|}\left(1-e^{2 \Im(\omega)(v-v_0)} \right)\right] \ ,
\end{equation}
where we have defined the integral $\mathcal{I}(r)$ as
\begin{align}\label{M-integral-1}
    &\mathcal{I}(r) \coloneqq  \int_{r_0}^{r}\dd \widetilde{r} \left[\mu^2 + f(\widetilde{r})|Z(\widetilde{r})|^2 \right] e^{-\mathcal{A}\widetilde{r}/r_0}\,  (\widetilde{r}/r_0)^{\mathcal{B} } |S(\widetilde{r})|^2  = \nonumber\\
    & \qquad =  \sum_{n \geq 0}  \Bigg\{ r_0 \sum_{p=0}^n\left(a_p \overline{a}_{n-p} \right)\left(\mu^2 I_{n}^{(\mathcal{A},\mathcal{B})}(r/r_0) + |\zeta|^2 I_{n+1}^{(\mathcal{A},\mathcal{B})}(r/r_0) + |\xi|^2 I_{n+1}^{(\mathcal{A},\mathcal{B}-2)}(r/r_0) + 2\Re(\zeta \overline{\xi}) I_{n+2}^{(\mathcal{A},\mathcal{B}-1)}(r/r_0)\right) + \nonumber \\
    & \qquad  \qquad  + \sum_{p=0}^n\bigg[  2(p+1)\Re(\overline{\zeta} a_{p+1}\overline{a}_{n-p}) I_{n+2}^{(\mathcal{A},\mathcal{B}-2)}(r/r_0) + 2(p+1)\Re(\overline{\xi} a_{p+1}\overline{a}_{n-p})I_{n+2}^{(\mathcal{A},\mathcal{B}-3)}(r/r_0) + \nonumber \\
    & \qquad \qquad \qquad + \frac{(p+1)(n-p+1) a_{p+1}\overline{a}_{n-p+1} }{r_0}I_{n+2}^{(\mathcal{A},\mathcal{B}-4)}(r/r_0)\bigg] \Bigg\} \ ,
\end{align}
where the constants $\zeta$ and $\xi$\,, which are used to define the auxiliary function $Z(r)$ in Eq.~\eqref{z-definition}, are reported in Eq.~\eqref{zeta-xi-def}. The integral function is defined as
\begin{equation}\label{Inth-def}
    I_n^{(\mathcal{A},\mathcal{B})}(r/r_0)\coloneqq\int_1^{r/r_0}\dd x \, e^{-\mathcal{A} x} x^{\mathcal{B}} \left(1 - \frac{1}{x} \right)^n  \ ,
\end{equation}
which can be analytically expressed as in Eq.~\eqref{I-nth}, or as derivatives of Euler's Gamma function as in Eq.~\eqref{integral-nth-power}.

\begin{figure}
\centering
    \includegraphics[width=\linewidth]{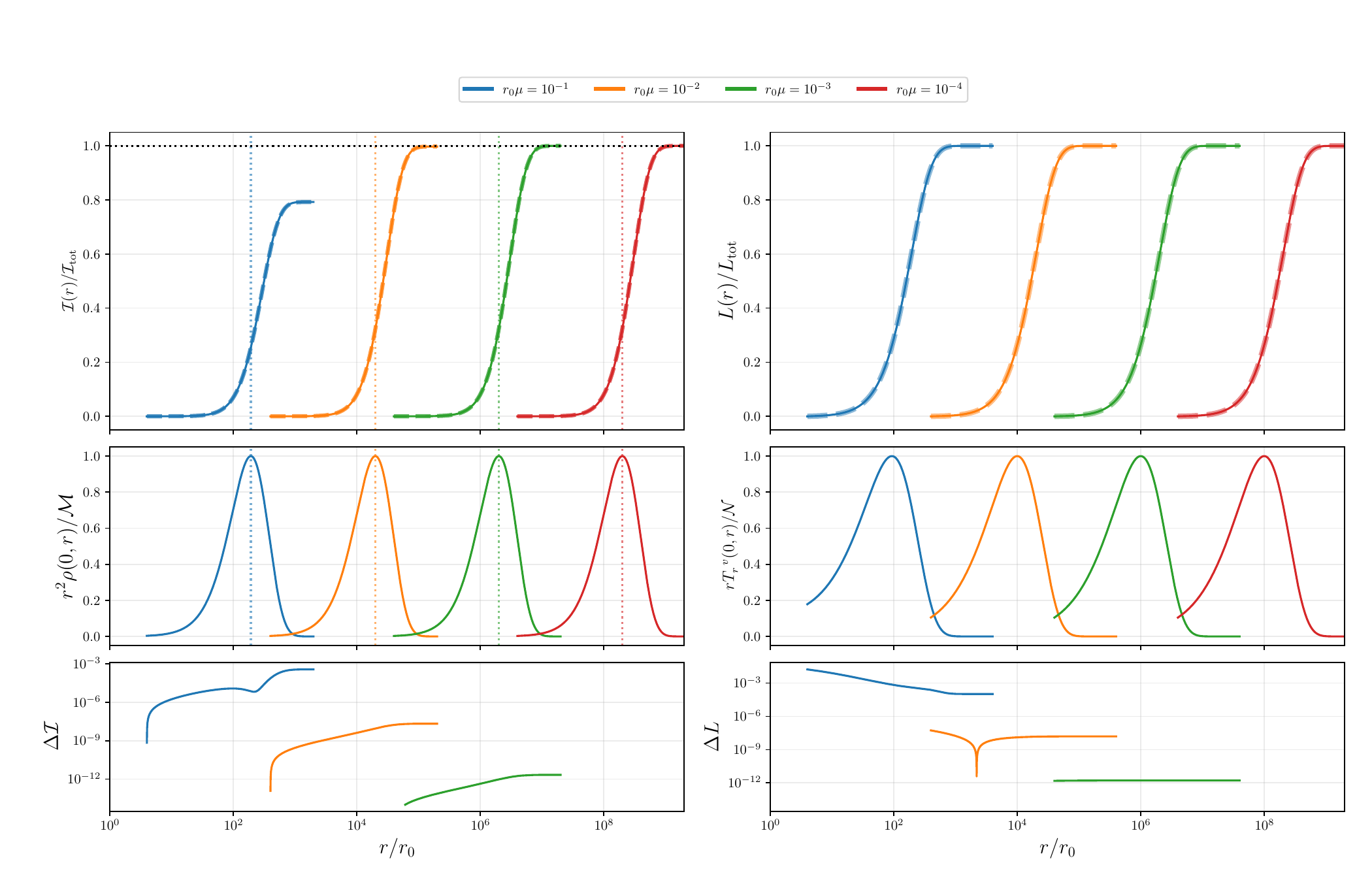}
    \caption{Top-left panel: the integral defined in Eq.~\eqref{M-integral-1}, normalized to its asymptotic value~\eqref{Itot-def}, is shown. The continuous line corresponds to the numerical evaluation of the integral appearing in the first line of Eq.~\eqref{M-integral-1}, while the dashed line represents the result obtained from the series of integrals given in the remaining terms of the same equation.
    Top-right panel: the integral function $L(r)$ obtained from the numerical integration of the last expression in Eq.~\eqref{L-integral-def} (continuous line) is compared with the result obtained from the series expansion~\eqref{1-over-e-lambda} (dashed line). Middle panels: the integrands appearing in Eq.~\eqref{M-integral-1} (left panel) and Eq.~\eqref{L-integral-def} (right panel) are shown. In the left panel, the dashed lines indicate the position of the maximum of the integrand, corresponding to the inflection point of the integrals displayed in the top-left panel. Bottom panels: the relative errors between the numerical evaluation of the integral representations and the corresponding series expansions are shown. Values below $10^{-14}$ are omitted, as they are dominated by floating-point numerical precision. All curves are computed for different values of $\rs\mu$, using the Leaver expansion~\eqref{S definition} truncated at order $N=10^3$.}
    \label{Fig-I-lambda}
\end{figure}
We can express the solution in Eq.~\eqref{lambda-sol} in a similar fashion as 
\begin{equation}
    e^{-\lambda(v,r)} = \frac{G \epsilon^2}{\pi} e^{2 \Im(\omega) v}L(r) + K(v) \ ,
\end{equation}
where we have defined the radial integral
\begin{equation}\label{L-integral-def}
    L(r/\rs) \coloneqq \int_{1}^{r/\rs}  \dd x \, |Z(x)|^2 e^{-\mathcal{A} x} \, x ^{\mathcal{B}-1} \, |S(x)|^2 \ .
\end{equation}
The details of the derivation and convergence of $\mathcal{I}(r)$ are discussed in the Appendix~\ref{Appintegrals} and, using the results in there, the integral in Eq.~\eqref{L-integral-def} can be expressed in a similar way to~\eqref{M-integral-1}
\begin{align}\label{1-over-e-lambda}
    L(r/\rs) &= \sum_{n \geq 0} \Bigg[ \sum_{p=0}^n\left(a_p \overline{a}_{n-p} \right) \left(|\zeta|^2 I_{n}^{(\mathcal{A},\mathcal{B}-1)}(r/\rs ) + |\xi|^2 I_{n}^{(\mathcal{A},\mathcal{B}-3)}(r/\rs) + 2 \Re(\zeta \overline{\xi}) I_{n}^{(\mathcal{A}, \mathcal{B} -2)}(r/\rs)\right) + \notag \\
    & + \left.\sum_{p=0}^n \left( \frac{(p+1)(n-p+1)}{\rs^2}a_{p+1} \overline{a}_{n-p+1} I_{n}^{(\mathcal{A}, \mathcal{B} - 5)}(r/\rs) + \frac{2(p+1)}{\rs}\Re(\overline{\zeta}a_{p+1} \overline{a}_{n-p}) I_{n}^{(\mathcal{A}, \mathcal{B} - 3)}(r/\rs) + \right. \right. \notag \\
    &  \qquad\qquad  +  \left. \frac{2(p+1)}{\rs}\Re(\overline{\xi}a_{p+1} \overline{a}_{n-p}) I_{n}^{(\mathcal{A}, \mathcal{B} - 4)}(r/\rs)\right) \Bigg] \ .
\end{align}

%\begin{align}\label{1-over-e-lambda}
    %e^{-\lambda(v,r)} &= \frac{\epsilon^2 G \,e^{2 \Im(\omega) v}}{\pi}\sum_{n \geq 0} \Bigg[ \sum_{p=0}^n\left(a_p \overline{a}_{n-p} \right) \left(|\zeta|^2 I_{n}^{(\mathcal{A},\mathcal{B}-1)}(r/\rs ) + |\xi|^2 I_{n}^{(\mathcal{A},\mathcal{B}-3)}(r/\rs) + 2 \Re(\zeta \overline{\xi}) I_{n}^{(\mathcal{A}, \mathcal{B} -2)}\right) + \notag \\
    %& + \left.\sum_{p=0}^n \left( \frac{(p+1)(n-p+1)}{\rs^2}a_{p+1} \overline{a}_{n-p+1} I_{n}^{(\mathcal{A}, \mathcal{B} - 5)} + \frac{2(p+1)}{\rs}\Re(\overline{\zeta}a_{p+1} \overline{a}_{n-p}) I_{n}^{(\mathcal{A}, \mathcal{B} - 3)} + \right. \right. \notag \\
    %&  \qquad\qquad  +  \left. \frac{2(p+1)}{\rs}\Re(\overline{\xi}a_{p+1} \overline{a}_{n-p}) I_{n}^{(\mathcal{A}, \mathcal{B} - 4)}\right) \Bigg] + K(v) \ .
%\end{align}
Finally, taking the $r\to+\infty$ limit of the continuity equation~\eqref{Flux-1} and using the expression \eqref{flux-r0} for the energy flux through the horizon, we obtain the identity
\begin{equation}\label{Eq:integrated_balance}
4 \pi \int_{r_0}^{+\infty} \widetilde{r}^2 \rho(v,\widetilde{r}) \, \dd \widetilde{r} = \epsilon^2 \frac{|\omega|^2e^{2 \Im(\omega)v-\mathcal{A}}}{2 \pi |\Im(\omega)|}~.
\end{equation}
When both sides are evaluated at $v=v_0$\,, this identity allows us identify the mass of the cloud in its initial configuration, computed on the initial slice at $v=v_0$\,, with the mass increase of the final black-hole state in the $v\to+\infty$ limit, after the scalar field has been completely absorbed by the hole. This interpretation is readily confirmed by a direct comparison with the solution for the Misner-Sharp mass \eqref{M-sol}. Using Eqs.~\eqref{Tvv}, \eqref{phi-modulus-squared}, \eqref{M-integral-1}, the identity \eqref{Eq:integrated_balance} can be re-expressed as 
\begin{equation}\label{Itot-analytic}
\mathcal{I}_{\rm tot}=\frac{|\omega|^2}{|\Im(\omega)|}e^{-\mathcal{A}}~,
\end{equation}
where we defined
\begin{equation}\label{Itot-def}
    \mathcal{I}_{\rm tot} \coloneqq \int_{r_0}^{\infty} \left[\mu^2 + f(\widetilde{r})|Z(\widetilde{r})|^2 \right] e^{-\mathcal{A}\widetilde{r}/r_0}\,  (\widetilde{r}/r_0)^{\mathcal{B} } |S(\widetilde{r})|^2 \, \dd \widetilde{r}  \ .
\end{equation}
In the top panels of Fig.~\ref{Fig-I-lambda}, we display the radial profiles of the two quantities entering our analysis, obtained by truncating Leaver's series~\eqref{S definition} at $N = 10^3$ coefficients, for different values of $\rs\mu$. In each panel, the continuous line shows the integral expression and the dashed line shows the corresponding series-of-integrals representation. The top-left panel shows the integral~\eqref{M-integral-1}, normalized to its analytical asymptotic value~\eqref{Itot-analytic}. As $\rs\mu$ decreases, the scalar field becomes more weakly bound and spreads further from the black hole, reflected in the outward shift of the inflection point of $\mathcal{I}(r)$ (i.e., the location of the maximum of the integrand~\eqref{M-integral-1}), shown normalized to its maximum value in the center-left panel. At large $r$, the integral saturates to $\mathcal{I}_{\rm tot}$ due to the exponential suppression of the integrand. The plot shows that the integral fails to saturate to the asymptotic value given in Eq.~\eqref{Itot-analytic} when $\rs \mu = 0.1$, suggesting that this case falls outside the regime of validity of the low-mass approximation, and the analytic formulae for the frequencies~\eqref{detweiler-re-im-omega-swave} are not suitable. The top-right panel shows the integral function $L(r)$, computed via the integral expression~\eqref{L-integral-def} and its series representation~\eqref{1-over-e-lambda}, for varying $\rs\mu$.  The center-right panel shows the integrand of Eq.~\eqref{lambda-sol} at the initial time $v = v_0 = 0$. 
The bottom panels show the relative errors between the two representations shown above, i.e., between the continuous and dashed lines in the top panels. The bottom-left panel shows the relative error between the integral form of $\mathcal{I}(r)$ and its series representation, both given in~\eqref{M-integral-1}. The bottom-right panel shows the relative error between the integral form of $L(r)$, given in~\eqref{L-integral-def}, and its series representation~\eqref{1-over-e-lambda}. In both cases the error remains small across the radial range considered, confirming the agreement between the two representations. As for the metric function in~\eqref{lambda-sol}, we can consider $K(v) = - \frac{G \epsilon^2}{\pi}e^{2 \Im(\omega) v}L_{\rm tot} + 1$ as a possible choice for the integration function such that we recover the asymptotic flatness of the spacetime
\begin{equation}
    e^{\lambda(v,r)} = \frac{1}{\frac{G \epsilon^2}{\pi}e^{2 \Im(\omega) v} (L(r)- L_{\rm tot}) + 1} \ ,
\end{equation}
where the asymptotic value $L_{\rm tot}$ is defined in analogy to $\mathcal{I}_{\rm tot}$~\eqref{Itot-def}, as $L_{\rm tot} \coloneqq \lim_{r \to \infty}L(r)$, so that when either $r \to \infty$ or $v \to + \infty$, we have $e^{\lambda} \to 1$ \ .
The parameter $\epsilon$ sets the size of the jump of $e^{\lambda}$ between its horizon value, $e^{\lambda(0,\rs)} $, and $1$, its value at spatial infinity. At fixed $\epsilon$, the shape and location of this transition are instead governed by $\rs\mu$ through the integrand in~\eqref{lambda-sol}.

\subsection{Trapping horizon and horizon mass}
The apparent horizon $r_H$ is implicitly determined by the horizon mass $M_H$ through the following equation
\begin{equation}\label{apparent-horizon-def}
    r_H(v) =  2 GM(v,r_H(v)) \ .
\end{equation}
Differentiating both sides of this identity with respect to $v$ gives
\begin{equation}
    \frac{\dd r_H}{\dd v} = 2 G \left( (\partial_v M)|_{r_H} + (\partial_r M)|_{r_H} \frac{\dd r_H}{\dd v}  \right) \ .
\end{equation}
Solving this equation algebraically for the horizon growth rate, we obtain
\begin{equation}\label{apparent-horizon-diffeq}
    \frac{\dd r_H}{\dd v} = \frac{2G\, \partial_v M_H}{1 - 2G \partial_r M_H} \ .
\end{equation}
Using equations~\eqref{BabichevMprime} and~\eqref{BabichevMdot}, we re-express the r.h.s.~in terms of the energy density and accretion flux~\eqref{rEF-flux-definition} 
\begin{equation}\label{accretion-rate}
    \dv{r_H}{v} = \frac{2G\, \mathcal{F}(v,r_H)}{1-8 \pi G \, r_H^2 \, \rho(v,r_H)} \ .
\end{equation}
Similar formulae for the apparent horizon have been derived earlier in \cite{Booth:2005ng,deCesare:2022aoe,deCesare:2023rmg} using different methods.
The expression for the accretion rate \eqref{accretion-rate} can be simplified considering an expansion in the test-field parameter
\begin{equation}
    r_H(v) = \rs + \delta r_H(v) \ ,
\end{equation}
and retaining only the terms of $\order{\epsilon^2}$ and using Eq.~\eqref{flux-r0}, we obtain
\begin{equation}
    \dv{(\delta r_H)}{v} =  2 G \,  \mathcal{F}(v,\rs) + \order{\epsilon^4} = \frac{2G\epsilon^2 |\omega|^2}{\pi}e^{2 \Im(\omega)v-\mathcal{A}}  + \order{\epsilon^4} \ ,
\end{equation}
which can be easily integrated with the condition that $r_H(v_0)= \rs$, obtaining
\begin{equation}\label{apparent-horizon-sol}
    r_H(v) = r_0 + \epsilon^2\frac{G |\omega|^2e^{2\Im(\omega)v_0-\mathcal{A}}}{ \pi \,|\Im(\omega)|}\left(1-e^{2 \Im(\omega) (v-v_0)}\right) + \order{\epsilon^4} \ .
\end{equation}
Dividing by $2G$ we obtain the apparent horizon mass, $M_H(v)=r_H(v)/(2G)$. This expression is formally analogous to the charging of a capacitor in an $RC$ circuit: the horizon grows monotonically from $r_0$ and saturates exponentially on the characteristic timescale $\tau = (2|\Im(\omega)|)^{-1}$, as the black hole progressively absorbs the surrounding scalar cloud.

To further characterize the geometric properties of the apparent horizon $r_H$, let us consider two future-pointing radial null vector fields $\mathfrak{l}$ and $\mathfrak{n}$ (respectively, outgoing and ingoing)
\begin{equation}
    \mathfrak{l} = \pdv{v} + \frac{e^{\lambda(v,r)}}{2}\left( 1 - \frac{2GM(v,r)}{r}\right)\pdv{r} \qq{and} \mathfrak{n} = -e^{-\lambda(v,r)} \pdv{r} \ ,
\end{equation}
normalized such that $\mathfrak{l}_\mu \mathfrak{n}^\mu= -1 $.
At the apparent horizon $r=r_H$, we have for the expansion scalars 
\begin{equation}
    \theta_\mathfrak{l} = \frac{e^{\lambda(v,r_H)}}{r_H}\left(1 - \frac{2GM(v,r_H)}{r_H} \right) = 0 \qq{and} \theta_\mathfrak{n} = - \frac{2 e^{-\lambda(v,r_H)}}{r_H} < 0 \ ,
\end{equation}
the latter condition is satisfied as long as the integration function in Eq.~\eqref{lambda-sol} is such that $K(v) > - 4 \pi G \int_{\rs}^{r_H} \widetilde{r}\,  \tensor{T}{_r^v}\, \dd \widetilde{r}$.
Then, $r_H$ is a future outer trapping horizon provided that the Lie derivative of $\theta_\mathfrak{l}$ along $\mathfrak{n}$ is negative
\begin{equation}
    \left.\pounds_\mathfrak{n} \theta_\mathfrak{l} \right|_{r=r_H} = \frac{2G\partial_r M_H-1}{r_H^2} < 0 \ .
\end{equation}
Using Eq.~\eqref{BabichevMprime}, we obtain that the above inequality is satisfied provided that
\begin{equation}\label{trapping-condition}
    8 \pi G r_H^2 \rho (v,r_H) < 1 \ .
\end{equation}
Expanding the l.h.s.~perturbatively and then differentiating w.r.t.~$v$, we obtain
\begin{equation}
    8\pi G \dv{v}\left[r_H(v)^2 \rho(v,r_H(v)) \right]= \frac{2 \epsilon^2 G \mu^2\Im(\omega)}{\pi} e^{2 \Im(\omega) v -\mathcal{A}} + \order{\epsilon^4} \ .
\end{equation}
Recalling that $\Im(\omega)<0$, the above equation implies that the term on the l.h.s.~of Eq.~\eqref{trapping-condition} is monotonically decreasing and the trapping condition can be inspected at the initial time $v=v_0$, which we take $v_0=0$ for simplicity, this condition becomes
\begin{equation}
    \frac{\epsilon^2 G\mu^2 e^{-\mathcal A}}{\pi} < 1 \ .
\end{equation}
Using Eq.~\eqref{curlyAdef} and within the low-mass approximation,
we have that $e^{-\mathcal{A}} \sim e^{-\rs^2 \mu^2} \approx 1 $, which allows to simplify the above condition as
\begin{equation}
    \epsilon \, \frac{\mu}{10^{-22}\,{\rm eV}}\lesssim 7.5 \cdot 10^{49} \ ,
\end{equation}
which is satisfied in the parameter range considered here, and allows us to conclude that $r_H$ is the future outer trapping horizon of the black hole, according to Hayward's definition~\cite{Hayward:1993wb}. The accretion rate~\eqref{accretion-rate} contains a non-linear dependence on $r_H$. However the trapping horizon condition~\eqref{trapping-condition} allows to conclude that the system never approaches the pole of~\eqref{accretion-rate} within the parameters in consideration.

Since the scalar field is exponentially localized, the Misner--Sharp mass
admits a well-defined asymptotic limit, equal to the total mass of the black-hole–scalar-field system,
$M_{\rm tot} \coloneqq \lim_{r\to \infty}M(v,r)$, which is a conserved quantity.
Computing the limit, we obtain
\begin{equation}\label{total-mass}
\begin{split}
    M_{\rm tot} \coloneqq \lim_{r\to \infty}M(v,r) &= M_0 +  \frac{\epsilon^2 \, }{2\pi} \left[e^{2 \Im(\omega) v}\,\mathcal{I}_{\rm tot} +  \frac{|\omega|^2e^{2 \Im(\omega)v_0-\mathcal{A}}}{|\Im(\omega)|}\left(1-e^{2 \Im(\omega)(v-v_0)} \right) \right] =\\
    &= M_0 + \frac{\epsilon^2 |\omega|^2 e^{2 \Im(\omega) v_0 - \mathcal{A}}}{2\pi |\Im(\omega)|}~.
\end{split}
\end{equation}
where in the last line we substituted the expression \eqref{Itot-analytic} for ${\cal I}_{\rm tot}$, thus making $M_{\rm tot}$ manifestly $v$-independent. Hence, we can define the total mass of the scalar (to the perturbative order being retained) as
\begin{equation}\label{cloud-mass-def}
    M_{\rm cloud}(v) \coloneqq M_{\rm tot} - M_H(v) = 4 \pi \int_{r_H(v)}^\infty r^2 \rho(v,r) \, \dd r  +\mathcal{O}(\epsilon^4)\ .
\end{equation}
The quantity $M_{\rm tot}$ is evaluated as the $r\rightarrow\infty$ limit of
the Misner--Sharp mass on hypersurfaces of constant advanced time $v$. Since
the scalar field decays exponentially at large distances, the energy flux
through null infinity vanishes, and therefore $M_{\rm tot}$ {is conserved by the dynamics}. 
In this setting, $M_{\rm tot}$ coincides with the total
mass of the black-hole--scalar-field system and provides a natural measure of
the total energy contained in the spacetime.
Differentiating Eq.\eqref{cloud-mass-def}, we obtain a detailed balance condition
\begin{equation}
    \dv{M_H}{v} + \dv{M_{\rm cloud}}{v} = 0 \ .
\end{equation}
Moreover, using Eqs.~\eqref{cloud-mass-def},~\eqref{Tvv} and~\eqref{phi-modulus-squared}, we obtain the following coupled ODE system, which satisfies detailed balance
\begin{equation}
    \dv{M_{\rm cloud}}{v} = 2 \Im(\omega) M_{\rm cloud}+\mathcal{O}(\epsilon^4) \qq{and} \dv{M_H}{v} = - 2 \Im(\omega) M_{\rm cloud}+\mathcal{O}(\epsilon^4) \ .
\end{equation}
Using the conservation of the total mass~\eqref{total-mass} we obtain the following long-time asymptotics for the system
\begin{equation}
    \lim_{v \to + \infty} M_{\rm cloud}(v) = 0 \qq{and} \lim_{v \to + \infty} M_H(v) = M_{\rm tot} \ .
\end{equation}
The total fractional growth of the black-hole mass upon complete absorption of the cloud is
\begin{equation}\label{fractionalgrowth}
    \lim_{v \to +\infty} \frac{M_H(v)-M_0}{M_0}=\lim_{v \to +\infty}\frac{r_H(v)-r_0}{r_0}=
    \epsilon^2\frac{G |\omega|^2e^{-\mathcal{A}+2\Im(\omega)v_0}}{\pi \, r_0 \, |\Im(\omega)|} = \epsilon^2\frac{4G}{ \pi r_0^2}\frac{e^{-(r_0 \mu)^2+ {\cal O}(\rs \mu)^6}}{(r_0 \mu)^4} \left(1 + {\cal O}(\rs \mu)^{2} \right)\ ,
\end{equation}
where the expression in the last step follows from the low-mass approximation~\eqref{detweiler-omega2-chi}. While the second factor diverges as $r_0\mu \to 0$\,, the prefactor $G/r_0^2 \sim (M_{\rm P}/M_0)^2$ is negligibly small for any astrophysically relevant black hole, ensuring that the total growth remains well controlled. The self-consistency of perturbation theory requires that $r_H(v)/r_0 -1 \ll 1$ at all times $v>v_0$\,. 
\begin{figure}
    \centering
\includegraphics[width=0.5\linewidth]{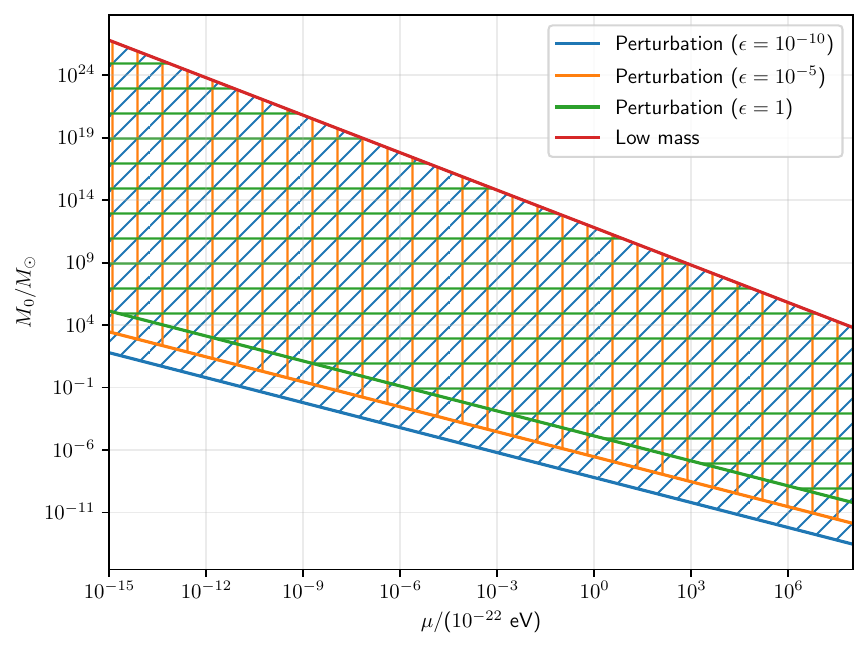}
    \caption{The colored regions denote the ranges of allowed initial black hole masses, expressed in units of $M_\odot$\,, as functions of the scalar field mass (normalized to $10^{-22}\,{\rm eV}$, which is a typical value for ultralight scalars \cite{Hu:2000ke,Hui:2021tkt}). Different colors correspond to different values of the field amplitude $\epsilon$\,. These regions correspond to the parameter space in which both the
    low-mass approximation for the dynamics of the test scalar field~\eqref{M0Detweilerbound} and the perturbative result for the backreaction on the geometry~\eqref{M0testfieldbound} are simultaneously valid.}
    \label{M0bounds}
\end{figure}
This condition, along with $\rs \mu\ll1$ defining the low-mass regime, implies
\begin{equation}\label{M0testfieldbound}
    \frac{4G}{ \pi r_0^2}\frac{\epsilon^2}{(r_0 \mu)^4} \ll 1 \implies M_0  \gg \epsilon^{1/3}\left(\frac{\mu}{10^{-22}~{\rm eV} }\right)^{-2/3}  1.37\cdot 10^{-5} M_\odot   \ ,
\end{equation}
which sets a lower bound on the initial black hole mass for the perturbative treatment to remain valid. Meanwhile, the low-mass condition for the scalar translates into the bound
\begin{equation}\label{M0Detweilerbound}
    M_0 \ll \left(\frac{\mu}{10^{-22}~{\rm eV} }\right)^{-1} 6.47 \cdot 10^{11} M_\odot \ . 
\end{equation}
In Fig.~\ref{M0bounds} we show the region of parameter space, spanned by the initial mass of the black hole and the mass of the scalar field, such that both inequalities \eqref{M0testfieldbound}, \eqref{M0Detweilerbound} are satisfied.
The plot shows that lighter scalar fields allow for larger values of the initial black-hole mass. The field amplitude $\epsilon$ controls the vertical shift of the lower bound of the allowed region. The evolution of the scalar field, the Misner-Sharp mass, and the trapping horizon is illustrated in Fig.~\ref{fig:evolution}.

\begin{figure}
\centering
\subfloat[]{\scalebox{0.55}{\includegraphics{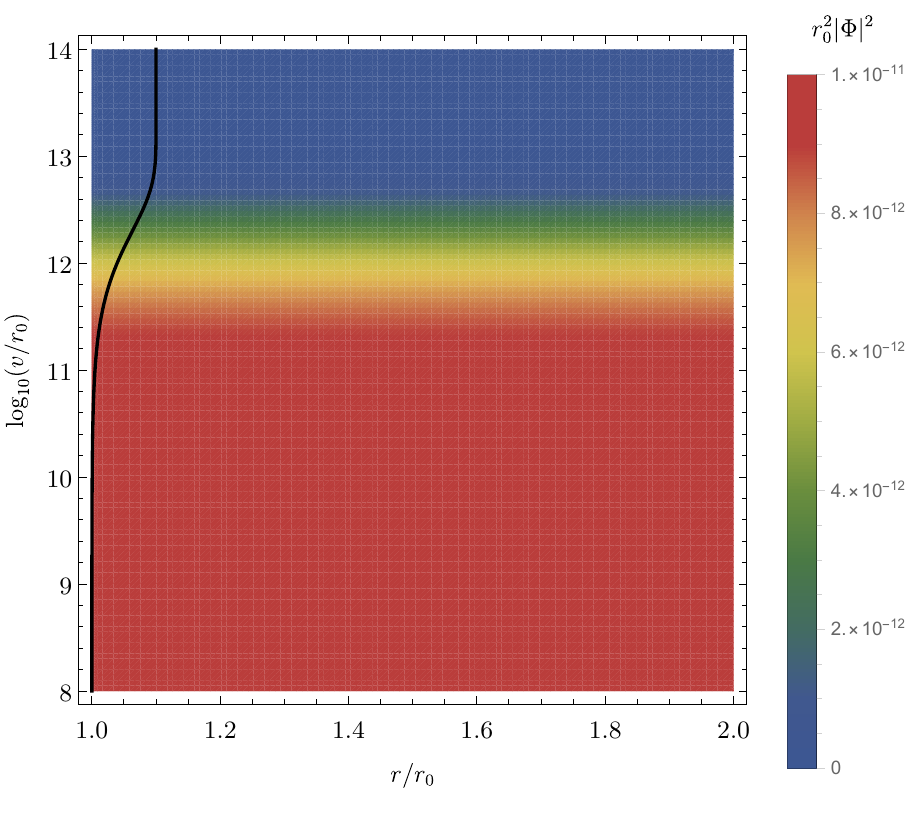}}}
\enspace
\subfloat[]{\scalebox{0.55}{\includegraphics{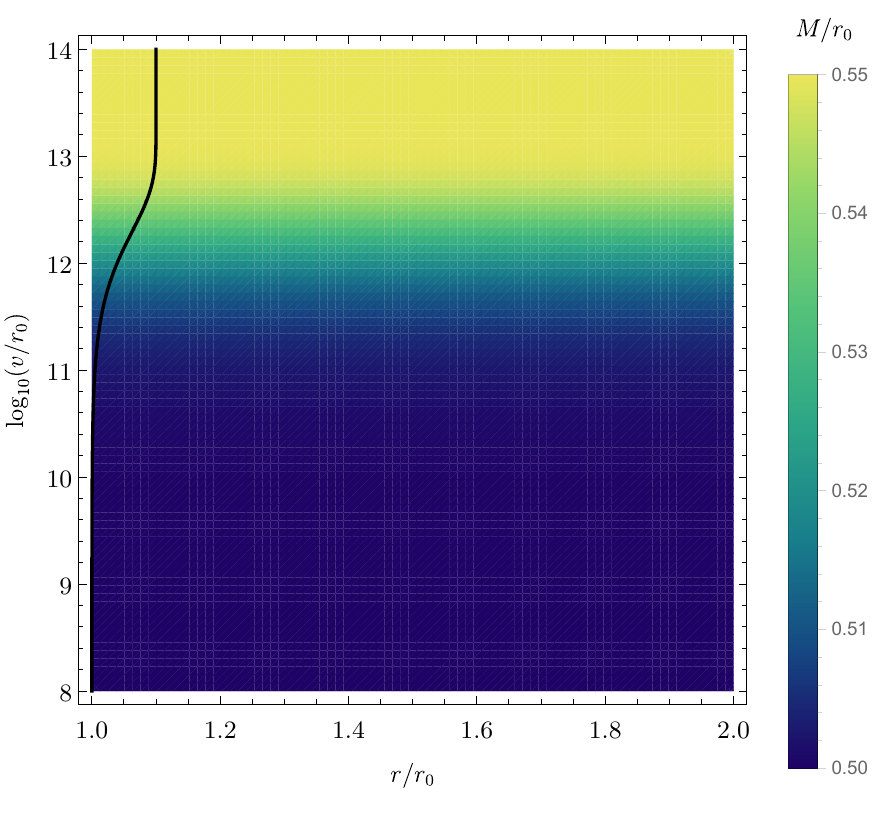}}}
\vspace{1mm}
\subfloat[]{\scalebox{0.55}{\includegraphics{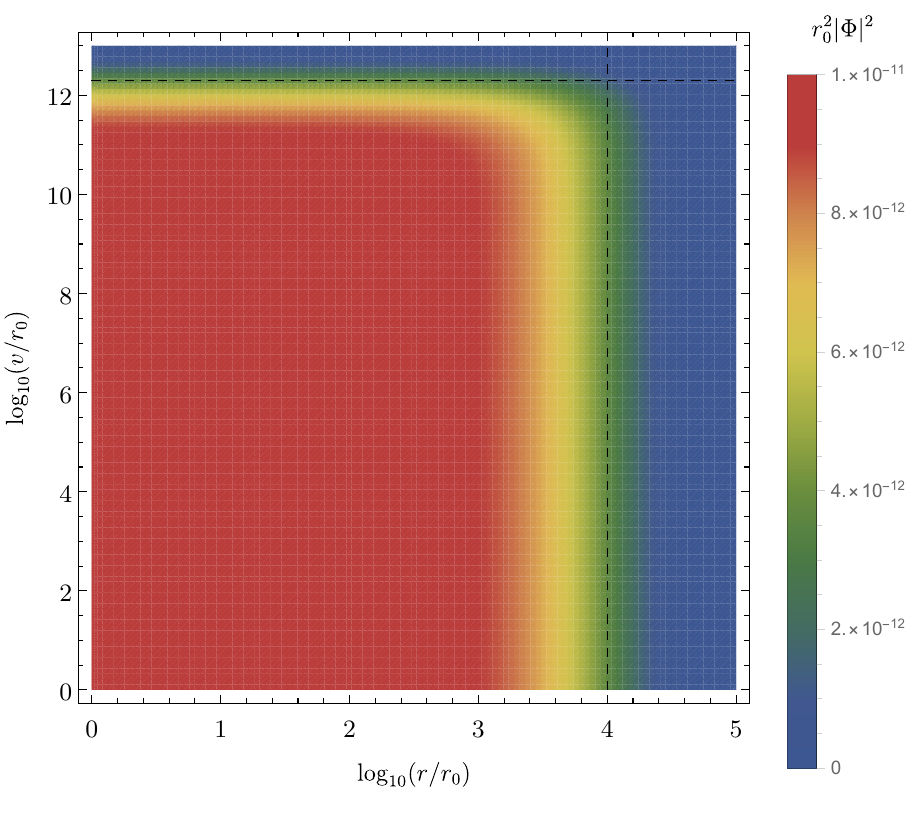}}}
\enspace
\subfloat[]{\scalebox{0.55}{\includegraphics{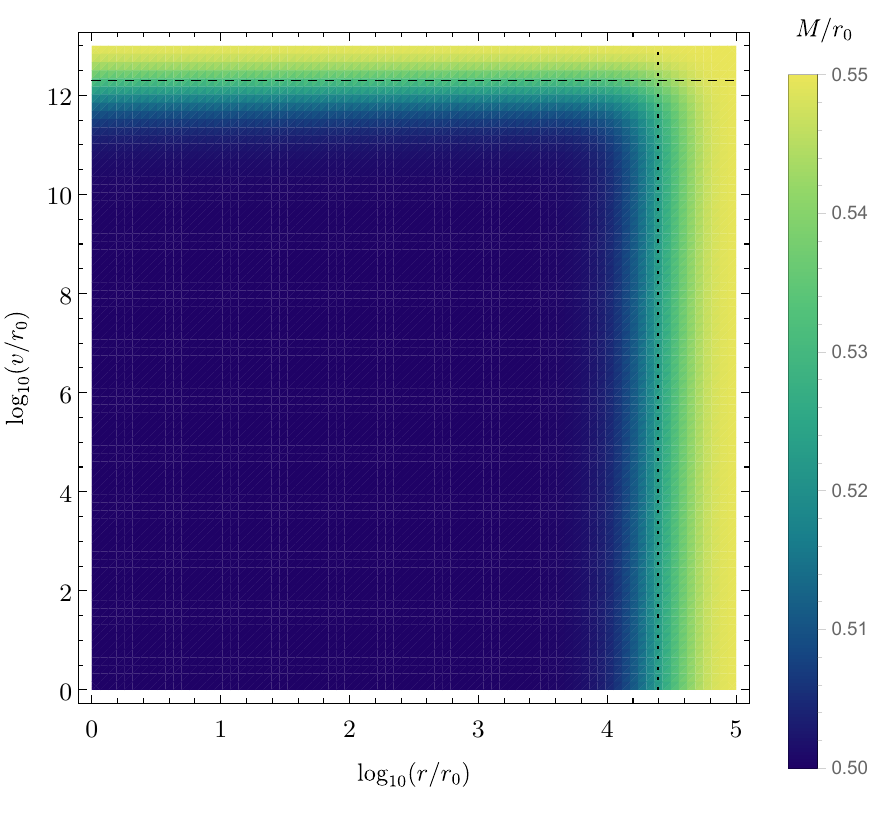}}}
\caption{The plots show the evolution of the scalar field `wig' configuration and the Misner-Sharp mass of the perturbed spacetime, with the following choice of parameters: the scalar field has mass $\mu=10^{-2}/r_0$ and amplitude $\epsilon \approx 2.803\times 10^{-5}$, such that the final horizon mass of the black hole after the scalar field has been completely absorbed is ten percent larger than the initial mass of the background geometry, i.e.~$M_{\rm final}=1.1 M_0$. The qualitative behaviour of the plots does not depend on the particular choice of parameters.
In the top panels, (a) and (b), the continuous black curve represents the trapping horizon, which expands as a result of the absorption of the scalar field from the black hole. In the bottom panels, a logarithmic scale is used on both axes, with a dashed horizontal line marking the characteristic decay time of the scalar wig $v_{c}=(2|{\rm Im}(\omega)|)^{-1}$. In panel (c) the dashed vertical line marks the characteristic decay radius of the scalar field $r_{c}=1/{\cal A}\approx 1/(r_0\mu)^2$, while in panel (d) the dotted vertical line  corresponds to the Misner-Sharp mass attaining $95\%$ of its asymptotic large-distance value, which occurs at a scale $\tilde{r}_c\sim 1/(r_0\mu^2)$\, up to an ${\cal O}(1)$ numerical prefactor.}
    \label{fig:evolution}
\end{figure}

Lastly, we compare our analysis with Ref.~\cite{Sanchis-Gual:2025lbp}, where the interaction between the black hole and the scalar wig is modeled as a coupled ODE system and studied numerically. The model uses Schwarzschild coordinates $(t,r)$\,, which are not regular at the black hole horizon, and this represents a major difference with our work. Further, in this reference the scalar field is assumed to evolve adiabatically as a function of $t$\,, tracking a sequence of quasi-stationary configurations on a slowly evolving black-hole background with instantaneous mass $M_{\rm BH}(t)$\,. The characteristic frequency and decay rate of the quasi-bound state, which are encoded in the complex frequency $\omega$ computed in the test-field approximation on a Schwarzschild background, are then promoted to a time-dependent quantity via $\omega = \omega(M_{\rm BH}(t))$\,. This yields a faster accretion rate compared to the perturbative result. However, it is worth stressing that the prescribed time dependence has not been derived from the Einstein equations, but is instead introduced through an ad-hoc prescription. 
In our approach, the time-dependence of the black-hole only arises as a backreaction effect, and  may only affect the scalar field dynamics at higher orders in perturbation theory. In principle, higher-order effects may lead to non-trivial effects, such as distortions of the scalar wig profile as computed in the test-field approximation, or the excitation of secondary harmonic components. Therefore, it is far from obvious that the resummation of higher-order perturbative corrections should result in a simple adiabatic time-dependence of the scalar frequency corresponding to the model proposed in Ref.~\cite{Sanchis-Gual:2025lbp}.

\section{Discussion}

We have presented an analytical treatment of the leading gravitational
backreaction of a spherically symmetric scalar wig on a Schwarzschild black
hole. Working in the small-coupling regime $r_0\mu\ll1$, we modelled the scalar
configuration by the fundamental quasi-bound $s$-wave solution of the
Klein--Gordon equation and computed its stress-energy tensor in the test-field
approximation. This source was then used to solve the Einstein equations
perturbatively to $\mathcal{O}(\epsilon^2)$ in ingoing
Eddington--Finkelstein coordinates, retaining the full time dependence of the
quasi-bound state rather than assuming a stationary matter distribution.

The resulting geometry is completely characterized by the metric functions
$M(v,r)$ and $\lambda(v,r)$, for which explicit analytic expressions were
obtained. From these solutions we derived the evolution of the apparent (future
outer trapping) horizon and its associated Misner--Sharp mass. The horizon
grows monotonically due to the influx of scalar energy and asymptotically
approaches a constant value on the characteristic timescale
$\tau=(2|\mathrm{Im}\,\omega|)^{-1}$, corresponding to the lifetime of the
quasi-bound state. At the same perturbative order, the mass stored in the
exterior scalar cloud decreases exponentially, and its loss is exactly balanced
by the increase of the horizon mass, consistently with the conservation of the
total asymptotic mass.

Our analysis extends previous perturbative studies of black-hole accretion,
which rely on a steady-state approximation for the matter distribution. Such an
approximation is appropriate only on timescales short compared to the lifetime
of the scalar cloud, during which the horizon flux is approximately constant.
In contrast, a scalar wig represents a finite bound reservoir of energy whose
stress-energy tensor decays exponentially in time. Retaining this time
dependence is therefore essential for describing the complete accretion process,
the depletion of the cloud, and the relaxation towards the final Schwarzschild
configuration. In this sense, the present work provides the time-dependent
generalization of steady-state accretion to the case of scalar quasi-bound
states.

The validity of our results relies on two independent assumptions. First, the
scalar field is described within a small mass approximation requiring
$r_0\mu\ll1$, which is well justified for ultralight scalars around black holes.
Second, the gravitational backreaction is treated perturbatively,
which requires the energy stored in the scalar cloud to remain small compared
with the black-hole mass. Within the overlap of these two regimes, illustrated
in Fig.~2, the perturbative solution provides a self-consistent description of
the coupled Einstein--Klein--Gordon system at leading order. Corrections to the
scalar profile induced by the evolving geometry enter only at higher orders in
the perturbative expansion and have not been considered here.

Although we focused on the fundamental spherically symmetric mode, the formalism
can be extended straightforwardly to higher radial overtones. In that case, the
structure of the backreaction equations remains unchanged, while the evolution
is controlled by the corresponding quasi-bound frequency and radial profile.
More generally, a superposition of quasi-bound states would produce several
characteristic decay times together with interference terms in the
stress-energy tensor. At sufficiently late times, however, the least damped
mode is expected to dominate the evolution.

An important extension of the present work concerns rotating black holes and
non-spherical scalar clouds. In that case, quasi-bound ``gravitational atom''
states with $\ell>0$ carry angular momentum and source both mass and angular
momentum accretion. Existing perturbative treatments of this problem still rely
on a steady-state approximation for the matter fields ~\cite{Kimura:2021dsa}.
Extending the present
time-dependent formalism to slowly rotating backgrounds would therefore provide
a natural framework for describing the complete backreacted evolution of
superradiant scalar clouds beyond steady state.

Finally, our analysis naturally suggests several future directions. It would be
interesting to compute the next perturbative order, allowing the scalar
configuration itself to respond to the evolving geometry, and to compare the
resulting evolution with fully nonlinear numerical simulations of the
Einstein--Klein--Gordon system. Another interesting avenue is to investigate the
competition between scalar accretion and Hawking evaporation for sufficiently
light black holes, although incorporating the latter consistently requires
going beyond the classical framework adopted in this work. 

\acknowledgments
We thank Flavio Rosales Infante for identifying a typo in an earlier version of the paper. This work is supported by the Center for Research and Development in Mathematics and Applications (CIDMA) (\url{https://ror.org/05pm2mw36}) under the Portuguese Foundation for Science and Technology (FCT -- Fundaç\~ao para a Ci\^encia e a Tecnologia, \url{https://ror.org/00snfqn58}), Grants UID/04106/2025 (\url{https://doi.org/10.54499/UID/04106/2025}) and UID/PRR/04106/2025 (\url{https://doi.org/10.54499/UID/PRR/04106/2025}), as well as the projects: Horizon Europe staff exchange (SE) programme HORIZON-MSCA2021-SE-01 Grant No.\ NewFunFiCO-101086251 and 2022.04560.PTDC (\url{https://doi.org/10.54499/2022.04560.PTDC}). MdC acknowledges support from INFN iniziativa specifica GeoSymQFT. The work of M. Del Piano is partially supported by the Carlsberg Foundation, grant CF22-0922. This work contributes to COST Action CA23130 -- Bridging high and low energies in search of quantum gravity (BridgeQG).

\appendix
\numberwithin{equation}{section}
\section{Quasi-resonant frequencies of a light scalar field}\label{AppDetweiler}

\subsection{Review of Detweiler's approximation}
Detweiler’s approximation~\cite{Detweiler:1980uk} yields analytic expressions for both the real and imaginary parts of the frequency of a massive scalar field with mass $\mu$ and multipole number $\ell \geq 1$ in a Kerr background of mass $M_0$ and spin parameter $a$. The approximation is valid in the regime where the Compton wavelength of the scalar field is much larger than the gravitational radius of the black hole, i.e. $G M_0 \mu \ll 1$\,. In this limit, the spectrum of the complex frequency can be expressed as

\begin{widetext}
\begin{subequations}\label{detweilers approx}
\begin{align}
     \Re(\omega_{\ell,n,m})  &= \mu \,  \sqrt{ 1 -\left( \frac{GM_0 \mu }{n+\ell +1 }\right)^2 } \ , \label{detweilers approx_real}\\
    \Im(\omega_{\ell,n,m}) & = \mu \,\left(GM_0 \mu\right)^{4 \ell +4}\left(\frac{ a \, m}{GM_0} - 2 \mu r_+\right)\frac{2^{4\ell +2}(2 \ell +1 +n)!}{(\ell +1 +n)^{2\ell+4}n!}\left[\frac{\ell!}{(2\ell)!(2\ell+1)!}\right]^2 \times \nonumber \\
    & \qquad \times \prod^\ell_{j=1} \left[ j^2 \left(1-\frac{a^2}{(GM_0)^2}\right)  + \left( \frac{ a \, m}{G M_0} - 2 \mu r_+\right) ^2\right] \ .\label{detweilers approx_imaginary}
\end{align}
\end{subequations}
\end{widetext}
Here $n$ and $m$ respectively label the overtones and the azimuthal component for a given value of the multipole moment $|m| \leq \ell$\,, while $r_+$ is the radius of the outer event horizon
\begin{equation}
    r_+ = GM_0\left(1+ \sqrt{1-\left(\frac{a}{G M_0}\right)^2} \right) \ .
\end{equation}
For non-spinning black hole, $a=0$\,, it reduces to $r_+ \equiv r_0 = 2 G M_0$\,.
Although the approximation is not intended to be valid for $\ell=0$\,, the authors of Ref.~\cite{Barranco:2012qs} argue that it is satisfactory agreement with both numerical and approximate expressions even in this scenario. However, the approximations and matching hierarchy used in
Ref.~\cite{Detweiler:1980uk} do not remain uniform in the $\ell=0$
Schwarzschild limit. The formal continuation of the final expression
therefore need not reproduce the correct leading coefficient for the
$s$-wave decay rate.
For this reason, the $s$-wave must be studied separately, which we do in the following with a careful implementation of matched asymptotic expansions. In particular, we will show that, while the imaginary part \eqref{detweilers approx_imaginary} extrapolated to the $a=\ell=0$ case predicts the correct mass scaling, its numerical prefactor does not match that of the $s$-wave mode in Schwarzschild. 

\subsection{Analytical determination of the quasi-resonant frequencies using matched asymptotic expansions}\label{sec:analytic-frequencies}
In Schwarzschild coordinates, the Klein-Gordon equation for the Fourier modes of the $s$-wave component ($\ell=0$) of the scalar field reads as
\begin{equation}\label{Eq:KleinGordon_swave}
\left(1-\frac{\rs}{r}\right)\dv[2]{\Phi_\omega}{r} +\frac{1}{r}\left(2-\frac{\rs}{r}\right)\dv{\Phi_\omega}{r}+\left(\frac{\omega^2}{1-\rs/r}-\mu^2\right)\Phi_\omega=0~.
\end{equation}
We introduce the following dimensionless combinations, $\tilde{\omega}\coloneqq r_0\omega$\,, $\tilde{\mu}\coloneqq r_0\mu$\,. It is convenient define a new radial coordinate $x=r/\rs -1$\,, and rescale the field as $\psi_\omega = x \,\Phi_\omega$\,.\footnote{A similar choice of rescaled variables in the Klein-Gordon equation was made in Ref.~\cite{deCesare:2025ccs}, for a massless scalar field in Schwarzschild-de Sitter.}
In terms of the new variables, the Klein-Gordon equation \eqref{Eq:KleinGordon_swave} reads as
\begin{equation}\label{Eq:KleinGordon_swave_xcoord}
x^2(1+x)\psi^{\prime\prime}(x)-x\psi^{\prime}(x)+\left( 1+ \tilde{\omega}^2(1+x)^3-\tilde{\mu}^2x(1+x)^2\right)\psi(x)=0~.
\end{equation}
(A prime is used to denote derivative with respect to $x$\,. The $\omega$ dependence is omitted, to make the notation lighter.)

We are interested in the regime where the mass and frequency of the scalar are both small and of the same order of magnitude, and the real part of the frequency is lower than the mass, so as to ensure an exponential decay in the large distance limit. Specifically, we assume $\tilde{\mu}\ll1$\,, $\Re(\tilde{\omega})\lesssim\tilde{\mu}$\,, and $\Re(\sqrt{\tilde{\mu}^2-\tilde{\omega}^2})\ll1$\,. We define a {\it near region} $\tilde{\mu}x\ll 1$\,, and a {\it far region} $x\gg 1$\,, whose overlap is guaranteed to exist due to the smallness of the dimensionless mass parameter $\tilde{\mu}$\,. In the following, we derive the asymptotic form of the solution in each region, and then match them in their overlap region $1\lesssim x\lesssim 1/\tilde{\mu}$\,.

\subsubsection{Near-region asymptotics}
In the near region, Eq.~\eqref{Eq:KleinGordon_swave_xcoord} can be approximated as
\begin{equation}\label{Eq:KleinGordon_swave_xcoord_near}
x^2(1+x)\psi^{\prime\prime}(x)-x\psi^{\prime}(x)+\left( 1+ \tilde{\omega}^2\right)\psi(x)=0~.
\end{equation}
The solution, subject to purely ingoing boundary conditions at the horizon, reads
\begin{equation}\label{Eq:nearregionsol}
\psi_{\rm near}(x)=C_{\rm near}x^{1-i\tilde{\omega}} {}_2F_1(1-i\tilde{\omega},-i\tilde{\omega},1-2i\tilde{\omega};-x)~.
\end{equation}
This has the following asymptotics
\begin{subequations}
\begin{align}
&\psi_{\rm near}(x)\approx C_{\rm near}x^{1-i\tilde{\omega}} ~,\quad \mbox{as}\; x\to 0~,\\
&\psi_{\rm near}(x)\approx C_{\rm near}(x+i\tilde{\omega})  ~,\quad \mbox{as}\; x\to +\infty~.\label{Eq:nearregionsol_farlimit}
\end{align} 
\end{subequations}

\subsubsection{Far-region asymptotics}
In the far region, Eq.~\eqref{Eq:KleinGordon_swave_xcoord} can be approximated as
\begin{equation}\label{Eq:KleinGordon_swave_xcoord_near}
\psi^{\prime\prime}(x)+\left(\tilde{\omega}^2-\tilde{\mu}^2+\frac{2\tilde{\omega}^2-\tilde{\mu}^2}{x} \right)\psi(x)=0~.
\end{equation}
We are interested in the solution that decays in the $x\to+\infty$ limit, which is
\begin{equation}
\psi_{\rm far}(x)=C_{\rm far} e^{-\sqrt{\tilde{\mu}^2-\tilde{\omega}^2} \, x}\, U\left(-k,0,2\sqrt{\tilde{\mu}^2-\tilde{\omega}^2} \, x\right)
~,\quad \mbox{with}\; k\coloneqq\frac{(2\tilde{\omega}^2-\tilde{\mu}^2)}{2\sqrt{\tilde{\mu}^2-\tilde{\omega}^2}}~,
\end{equation}
where $U$ denotes the confluent hypergeometric function. In the large distance limit, we have the asymptotics
\begin{equation}
    \psi_{\rm far}(x)\approx C_{\rm far}e^{-\sqrt{\tilde{\mu}^2-\tilde{\omega}^2} \, x} \left(2\sqrt{\tilde{\mu}^2-\tilde{\omega}^2} \, x\right)^k  ~,\quad \mbox{as}\; x\to +\infty~.
\end{equation}
For generic values of the parameter $k$, the asymptotic expansion of this solution in the $x\to0$ limit contains logarithmic terms, which cannot be matched to the near solution. Such logarithmic terms are however suppressed if $k$ is close to non-negative integer values. This gives a quantization condition for the quasi-resonant frequencies
\begin{equation}\label{Eq:quantization_condition}
    k=n+1+\delta~,
\end{equation}
with $n$ a non-negative integer and a small complex correction $|\delta|\ll1$. Assuming Eq.~\eqref{Eq:quantization_condition}, we obtain the following small-distance asymptotics for the far-region solution
\begin{equation}\label{Eq:farregionsol_nearlimit}
    \psi_{\rm far}(x)\approx  C_{\rm far} (-1)^{(n+1)}n! \left(\delta - 2\sqrt{\tilde{\mu}^2-\tilde{\omega}^2} (n+1)\,x  \right) ~,\quad \mbox{as}\; x\to 0~.
\end{equation}
\subsubsection{Matching conditions and determination of the spectrum}
Matching the large-distance asymptotics of the near-region solution \eqref{Eq:nearregionsol_farlimit} with the small-distance asymptotics of the far-region solution \eqref{Eq:farregionsol_nearlimit} in their overlap region, we obtain the following matching conditions
\begin{subequations}
    \begin{align}
        C_{\rm near}&=2(-1)^n (n+1)!\sqrt{\tilde{\mu}^2-\tilde{\omega}^2}\,C_{\rm far} ~,\label{Eq:matching_amplitudes}\\
        \delta&=-2i(n+1)\tilde{\omega}\sqrt{\tilde{\mu}^2-\tilde{\omega}^2}~.\label{Eq:matching_delta}
    \end{align}
\end{subequations}
Combining Eqs.~\eqref{Eq:quantization_condition} and \eqref{Eq:matching_delta}, we obtain an algebraic equation for $\tilde{\omega}$, which can be solved using standard analytical approximation methods
\begin{equation}
\tilde{\omega}=\tilde{\mu}\sqrt{1-\frac{\tilde{\mu}^2}{4(n+1)^2}}-i\frac{\tilde{\mu}^6}{4(n+1)^3}+{\cal O}(\tilde{\mu}^7)~.
\end{equation}
In terms of the original dimensionful variables, the quasi-resonant frequency spectrum is expressed as
\begin{equation}\label{Eq:spectrum_mae}
\omega\approx\mu\left(\sqrt{1-\frac{(r_0\mu)^2}{4(n+1)^2}}-i\frac{r_0^5\mu^5}{4(n+1)^3}\right)~.
\end{equation}

Comparing the spectrum \eqref{Eq:spectrum_mae} with the extrapolation of Detweiler's formula to the case $a=\ell=0$, we observe that the real part \eqref{detweilers approx_real} is in good agreement with our result, while for the imaginary part \eqref{detweilers approx_imaginary} gives $\Im\omega=-r_0^5\mu^6/2(1+n)^3$, which is off by a factor of 1/2 from the correct result just derived. In the remainder of this section, we compare the spectrum \eqref{Eq:spectrum_mae} with the numerical computation, to further corroborate our analytical results.

\subsection{Numerical determination of the quasi-resonant frequencies using Leaver's method}\label{sec:numerical-Leaver}

We determine the quasi-bound state frequencies numerically using Leaver's continued-fraction method, implemented in arbitrary precision (\texttt{mpmath}~\cite{mpmath}). The frequencies $\omega$ satisfy
\begin{equation}\label{Leaver_CF}
     0 = \beta_0 - \frac{\alpha_0 \gamma_1}{\beta_1 \, -}\frac{\alpha_1 \gamma_2}{\beta_2 - }\frac{\alpha_2 \gamma_3}{\beta_3 - \ldots} \ ,
\end{equation}
where the coefficients $\alpha_n$, $\beta_n$, $\gamma_n$ are given in Eqs.~\eqref{leaver-coefficients}. In practice the infinite continued fraction must be truncated at some large order $N$, and we follow Nollert's prescription~\cite{Nollert1993} of replacing the remainder by its large-$N$ asymptotics. Writing the remainder as
\begin{equation}\label{Nollert-implementation}
    R_N = \frac{\gamma_N}{\beta_N - \alpha_N R_{N+1}} \ , \qquad\qquad
    R_{N} = C_0 + \frac{C_1}{\sqrt{N}} + \frac{C_2}{N} + \order{N^{-3/2}}~,
\end{equation}
and matching order by order in powers of $N^{-1/2}$, we obtain
\begin{subequations}\label{Eq:nollert-coefficients}
\begin{align}
    C_0 &= -1 \ , \\
    C_1 &= \sqrt{2 \rs \chi} \ , \\
    C_2 &= \frac{3}{4} - \frac{3}{2}\,\rs \chi + \frac{\rs \omega^2}{2 \chi} \ . \label{Eq:C2}
\end{align}
\end{subequations}
The matching condition at order $N^{-3/2}$ is linear in $C_2$, so that this coefficient is determined uniquely and carries no branch ambiguity.

One feature of the small-coupling regime requires care. The asymptotic series is not ordered at the truncation orders that are numerically accessible. The next coefficient behaves as $C_3 \sim 3/(32\,\rs\mu)$ at small masses, so that $|C_3|N^{-3/2}<|C_2|N^{-1}$ only for $N\gtrsim3\times10^{-3}/(\rs\mu)^2$, that is $N\gtrsim10^{9}$ already at $\rs\mu=10^{-6}$, and $N\gtrsim10^{17}$ at $\rs\mu=10^{-10}$. Including further terms of the expansion therefore does not improve the result.

Accordingly, each frequency is evaluated at truncation orders $N$ and $2N$, and the leading truncation error, which is found to scale as $1/N$, is removed by Richardson extrapolation~\cite{richardson1911}
\begin{equation}\label{Eq:richardson}
  \omega_{\rm ext}=2\,\omega(2N)-\omega(N)~,
\end{equation}
the difference $|\Im\,\omega(2N)-\Im\,\omega(N)|$ being retained as an estimate of the residual truncation error. We stress that the residual of the root finding at fixed $N$ is not a measure of this error: it only shows that the root of the \emph{truncated} continued fraction has been located, and it remains below $10^{-50}$ even when the truncation error is of order $10^{-5}$.

The root-finder is seeded with points close to the analytic approximation~\eqref{Eq:spectrum_mae} for $n=0$, with displacements of $\order{10^{-6}}$, and a solution is accepted only if the residual of Eq.~\eqref{Leaver_CF} lies below a prescribed tolerance \emph{and} the resulting frequency satisfies $0<\Re(\omega)<\mu$ and $\Im(\omega)<0$, i.e.\ corresponds to a genuinely decaying bound state; this criterion excludes spurious roots and overtones. Frequencies are obtained via a continuation scheme in $\mu$: starting from the analytic seed at $\mu_{\rm start}$, each converged solution seeds the next value of $\mu$ through a power-law extrapolation based on the known small-$\mu$ scaling ($\Re(\omega)\propto\mu$, $\Im(\omega)\propto\mu^6$), which keeps the root-finder locked onto the fundamental mode throughout the scan. Both the truncation order $N$ and the working precision are increased adaptively as $\mu$ decreases, to control the catastrophic cancellation that otherwise degrades the continued-fraction evaluation at small $\mu$; on failure, the seed is perturbed and the step retried with refined $N$ and precision

As a convergence check, at $\rs\mu=10^{-6}$ and $N=4000$ the coefficient $|\Im(\rs\omega)|/(\rs\mu)^6$ equals $0.2499375$ without extrapolation and $0.2500000$ with it, to be compared with the exact small-mass limit $1/4$: the residual error is reduced by more than three orders of magnitude. Moreover, once extrapolated, the result is insensitive to the form adopted for the tail at the level of $5\times10^{-8}$, against $6\times10^{-5}$ without extrapolation. The frequencies used throughout this work are the extrapolated ones.

\begin{figure}
    \centering
    \includegraphics[width=\linewidth]{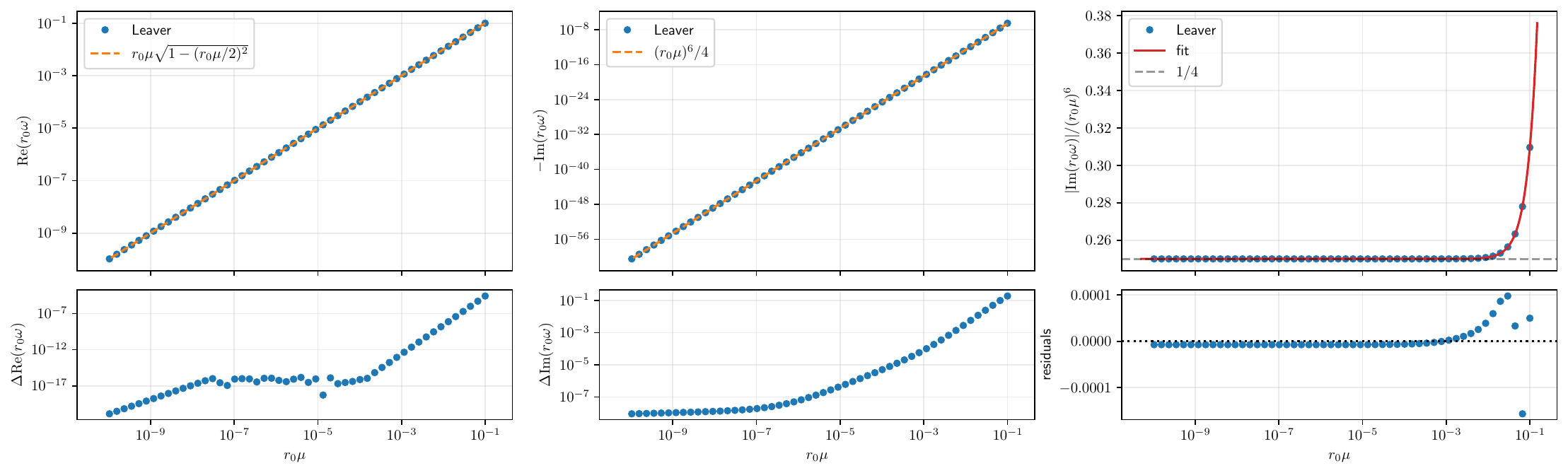}
    \caption{The top-left and top-centre panels show respectively the real and imaginary part of the fundamental mode frequency computed with Leaver's method (blue dots) and the analytic expressions~\eqref{Eq:spectrum_mae} (dashed orange lines), while the bottom panels show the relative errors between the numerical results and the analytic approximations. The top-right panel shows the imaginary part normalized by $(\rs \mu)^6$, fitted to the curve $c_0 + c_1 (\rs\mu)^p$ over the range $10^{-10}\leq\rs\mu\leq10^{-1}$. The best-fit parameters are $c_0 = 0.2500073 \pm 4.8 \times 10^{-6}$, $c_1 = 4.2068 \pm 2.1 \times 10^{-2}$, and $p = 1.8493 \pm 2.0 \times 10^{-3}$, with $\chi^2/{\rm d.o.f.} = 1.09\times10^{-9}$; holding $c_0$ at its analytic value $1/4$ gives $c_1 = 4.2000 \pm 2.1\times10^{-2}$ and $p = 1.8485 \pm 2.0\times10^{-3}$, with essentially the same residuals. The bottom-right panel shows the residuals of the fitting model relative to the data}
    \label{fig:Leaver_frequencies}
\end{figure}

Figure~\ref{fig:Leaver_frequencies} compares the frequencies obtained from this procedure with the analytic approximations derived in Sec.~\ref{sec:analytic-frequencies}. The agreement is excellent across the sampled range of $\mu$, with relative errors shown in the bottom panels. In particular, the fitted value of $c_0$ agrees with the analytic coefficient $1/4$ to three parts in $10^5$. Its residual offset is a property of the two-term fitting function rather than of the numerics: lowering the upper end of the fitting range drives $c_0$ monotonically onto $1/4$, through $0.2500013$ for $\rs\mu\leq3\times10^{-2}$, $0.2500005$ for $\rs\mu\leq10^{-2}$ and $0.2500000$ for $\rs\mu\leq10^{-4}$, while a fit restricted to $\rs\mu<10^{-5}$ returns $0.2500000030$. Independently of any fit, the computed coefficient at $\rs\mu=10^{-10}$ equals $0.250000002217$, and approaches $1/4$ monotonically from above.

To document the stability of the procedure, Table~\ref{tab:richardson-stability} compares the extrapolated results obtained from the pairs $(N,2N)$ with $N=10^3$ and $N=2\times10^3$, at four representative small masses. The two determinations agree to better than $4\times10^{-8}$ for $\rs\mu\leq10^{-6}$, and the residual deviation from the exact limit $1/4$ is reduced by about a factor of four when $N$ is doubled, consistent with an $\order{N^{-2}}$ remainder after the leading $1/N$ error has been removed. At $\rs\mu=10^{-4}$ the two extrapolations differ by $5\times10^{-7}$, a larger spread which is expected since the physical correction $c_1(\rs\mu)^p$ is itself of order $2\times10^{-7}$ at that mass and is therefore not negligible on this scale.

\begin{table}
\centering
\begin{tabular}{c|c|c|c|c}
\toprule
$\rs\mu$ & no extrapolation & ext.\ from $(N,2N)$ & ext.\ from $(2N,4N)$ & difference \\
\midrule
$10^{-4}$  & $0.249939681$ & $0.250001863$ & $0.250001327$ & $5.4\times10^{-7}$ \\
$10^{-6}$  & $0.249937516$ & $0.250000058$ & $0.250000023$ & $3.6\times10^{-8}$ \\
$10^{-8}$  & $0.249937495$ & $0.250000041$ & $0.250000010$ & $3.1\times10^{-8}$ \\
$10^{-10}$ & $0.249937495$ & $0.250000041$ & $0.250000010$ & $3.1\times10^{-8}$ \\
\bottomrule
\end{tabular}
\caption{Stability of the Richardson extrapolation~\eqref{Eq:richardson}. For each value of $\rs\mu$ we report the coefficient $|\Im(\rs\omega)|/(\rs\mu)^6$ evaluated at truncation order $N=4\times10^3$ without extrapolation, and the values extrapolated from two different pairs of truncation orders. The exact small-mass limit of the coefficient is $1/4$.}
\label{tab:richardson-stability}
\end{table}

\section{Computing the integrals in the mass function}\label{Appintegrals}
In section~\ref{sec-mass-function}, we have presented in Eq.~\eqref{M-sol-I} the mass function that solves the field equations~\eqref{Eq:efe} in terms of the integral~\eqref{M-integral-1}. In this section, we show the detailed computation of $\mathcal{I}(r)$\,. Let us start by writing out the $\tensor{T}{_v^v}$ component using Eq.~\eqref{Tvv} and~\eqref{phi-modulus-squared} and introduce the dimensionless integration variable $x \coloneqq r/r_0$\,, so we have
\begin{equation}
4\pi\int_{r_0}^r \dd r \, r^2 \rho(v,r) = \frac{\epsilon^2 \, e^{2 \Im(\omega) v}}{2 \pi}\,\mathcal{I}(r) \qq{with} \mathcal{I}(r) \coloneqq r_0\int_1^{r/r_0} \dd x \left[\mu^2 + f(x)|Z(x)|^2 \right] e^{-\mathcal{A}x}\,  x^{\mathcal{B} } |S(x)|^2  \ .
\end{equation}
where $\mathcal{A}$ and $\mathcal{B}$ are given in Eq.~\eqref{curlyAdef} and~\eqref{curlyBdef}, the auxiliary function $Z(x)$ is given in Eq.~\eqref{z-definition}, while $f(x) = 1 - 1/x$ is introduced as a shorthand notation. Using the Cauchy product, the quantity $|S(x)|^2$ can be expressed as
\begin{equation}
    |S(x)|^2 = 1 +\sum_{n \geq 1} \sum_{p=0}^n a_p \overline{a}_{n-p} f(x)^n \ ,
\end{equation}
where the overline indicates the complex conjugate and $a_0=1$\,. The integral is convergent if $\mathcal{A}$ is positive, and using \eqref{detweiler-re-im-omega-swave} and \eqref{detweiler-chi}, we have
\begin{equation}
    \mathcal{A} = (r_0\mu)^2 \left[1 - (r_0\mu)^4 + \order{(r_0 \mu)^5}\right] \ ,
\end{equation}
which is positive as long as the low-mass condition
$r_0\mu  \ll 1$ holds.
The squared norm of the auxiliary function $Z(x)$ can be expanded as
\begin{equation}\label{Z-norm-squared}
    |Z(x)|^2 = |\zeta|^2 + \frac{|\xi|^2}{x^{2}} + \frac{|S'(x)|^2}{r_0^2 \, |S(x)|^2} + \frac{2 \Re(\zeta \overline{\xi})}{x} + \frac{2}{r_0} \Re \left(\overline{\zeta}\frac{S'(x)}{S(x)} \right) + \frac{2}{r_0 \, x} \Re \left(\overline{\xi}\frac{S'(x)}{S(x)} \right) \ , 
\end{equation}
where $\zeta$ and $\xi$ are constants defined in~\eqref{zeta-xi-def}. We can express the integral in terms in different contributions that we report in the following bullet points:
\begin{itemize}
    \item The first contribution is proportional to the field mass squared
    \begin{equation}\label{integral-mu2-piece}
        \mu^2\int_1^{r/r_0} \dd x \, e^{-\mathcal{A} x}x^\mathcal{B} |S(x)|^2 = \mu^2 \int_1^{r/r_0}\dd x \left(\sum_{n \geq 0} \sum_{p=0}^n a_p \overline{a}_{n-p}\right) e^{-\mathcal{A} x}x^\mathcal{B} f(x)^n = \mu^2 \sum_{n \geq 0} \sum_{p=0}^n a_p \overline{a}_{n-p} I_n^{(\mathcal{A},\mathcal{B})}(r/r_0) \ , 
    \end{equation}
    where we have defined the following integral functions
    \begin{align}
        I_n^{(\mathcal{A},\mathcal{B})}(x) &\coloneqq \int_1^x \dd x \, e^{-\mathcal{A} x}x^\mathcal{B} f(x)^n = e^{-\mathcal{A}}\int_1^x \dd x \, e^{-\mathcal{A} (x-1)}x^{\mathcal{B}-n} (x-1)^n =  \notag \\
        & = (-1)^n e^{-\mathcal{A}}\dv[n]{\mathcal{A}}\left[e^{\mathcal{A}}\mathcal{A}^{n-\mathcal{B}-1}  \left( \gamma(\mathcal{B}-n+1, \mathcal{A}x ) - \gamma(\mathcal{B}-n+1,\mathcal{A}) \right) \right]\ ,\label{integral-nth-power}
    \end{align}
    where $\gamma(a+1,x)$ is the incomplete lower gamma function
    \begin{equation}\label{gamma-lower-def}
        \gamma(a+1,x) \coloneqq \int_{0}^{x}  e^{-t}t^{a}\,\dd t \qq{with} a \in \mathbb{R} \ .
    \end{equation} 
    The $n=0$ integral can be readily expressed as
    \begin{align}
        I_0^{(\mathcal{A},\mathcal{B})}(x) = \frac{\gamma(\mathcal{B}+1,\mathcal{A}x)-\gamma(\mathcal{B}+1,\mathcal{A})}{\mathcal{A}^{\mathcal{B}+1}} \ ,
    \end{align}
    and a full general expression of~\eqref{integral-nth-power} can be found in Eq.~\eqref{I-nth}.
    \item The following contributions are given in terms of power-law behaviors in Eq.~\eqref{Z-norm-squared}
    \begin{subequations}
        \begin{align}
        |\zeta|^2\int_1^{r/r_0} \dd x \, e^{-\mathcal{A} x}x^\mathcal{B} f(x) |S(x)|^2 &=  |\zeta|^2\sum_{n \geq 0} \sum_{p=0}^n a_p \overline{a}_{n-p} I_{n+1}^{(\mathcal{A},\mathcal{B})}(r/r_0) \ ,\\
        |\xi|^2\int_1^{r/r_0} \dd x \, e^{-\mathcal{A} x}x^{\mathcal{B}-2}f(x) |S(x)|^2 &=  |\xi|^2\sum_{n \geq 0} \sum_{p=0}^n a_p \overline{a}_{n-p} I_{n+1}^{(\mathcal{A},\mathcal{B}-2)}(r/r_0) \ ,\\
        2\Re(\zeta \overline{\xi})\int_1^{r/r_0} \dd x \, e^{-\mathcal{A} x}x^{\mathcal{B}-1} f(x) |S(x)|^2 &=  2\Re(\zeta \overline{\xi})\sum_{n \geq 0} \sum_{p=0}^n a_p \overline{a}_{n-p} I_{n+1}^{(\mathcal{A},\mathcal{B}-1)}(r/r_0) \ .
    \end{align}
    \end{subequations}
    \item Finally, one needs to take care of the terms involving $S(x)$ and $S^\prime(x)$\,, which can be handled first noting that
    \begin{equation}
        S'(x) = \frac{1}{x^2}\sum_{n\geq 0}(n+1)a_{n+1}f(x)^n \qq{hence} |S'(x)|^2 = \frac{1}{x^4}\sum_{n\geq 0}\left(\sum_{p=0}^n(p+1)(n-p+1)a_{p+1}\overline{a}_{n-p+1}\right)f(x)^n \ ,
    \end{equation}
    so that the mixed terms in the integral can be treated as
    \begin{multline}
        2 \Re \left(\overline{\zeta}\frac{S'(x)}{S(x)} \right) |S(x)|^2 = 2\Re(\overline{\zeta}S(x)\overline{S'}(x)) = \overline{\zeta}S(x)\overline{S'}(x)+ \zeta S'(x)\overline{S}(x) =\\= \frac{1}{x^2}\sum_{n\geq 0} \left[\sum_{p=0}^n (p+1)\left(\overline{\zeta}\, a_{p+1} \overline{a}_{n-p} + \zeta\,\overline{a}_{p+1}a_{n-p}\right)\right]f(x)^n = \frac{2}{x^2}\sum_{n\geq 0} \left[\sum_{p=0}^n (p+1)\Re\left(\overline{\zeta}\, a_{p+1} \overline{a}_{n-p}\right)\right]f(x)^n \ .
    \end{multline}
    Hence, the last three contributions are
    \begin{subequations}
        \begin{align}
            \frac{1}{r_0^2} \int_1^{r/r_0} \dd x \, e^{-\mathcal{A}x}x^{\mathcal{B}}f(x) |S^\prime(x)|^2 &= \frac{1}{r_0^2}\sum_{n\geq 0}\left(\sum_{p=0}^n(p+1)(n-p+1)a_{p+1}\overline{a}_{n-p+1}\right) I_{n+1}^{(\mathcal{A},\mathcal{B}-4)}(r/r_0) \ ,\\
            \frac{2}{r_0} \int_1^{r/r_0} \dd x \, e^{-\mathcal{A}x}x^{\mathcal{B}}f(x) \Re\left(\overline{\zeta} S^\prime(x) \overline{S}(x) \right) &= \frac{2}{r_0}\sum_{n\geq 0}\left(\sum_{p=0}^n(p+1)\Re\left( \overline{\zeta} a_{p+1}\overline{a}_{n-p}\right)\right) I_{n+1}^{(\mathcal{A},\mathcal{B}-2)}(r/r_0) \ ,\\
            \frac{2}{r_0} \int_1^{r/r_0} \dd x \, e^{-\mathcal{A}x}x^{\mathcal{B}-1}f(x) \Re\left(\overline{\xi} S^\prime(x) \overline{S}(x) \right) &= \frac{2}{r_0}\sum_{n\geq 0}\left(\sum_{p=0}^n(p+1)\Re\left( \overline{\xi} a_{p+1}\overline{a}_{n-p}\right)\right) I_{n+1}^{(\mathcal{A},\mathcal{B}-3)}(r/r_0) \ ,
        \end{align}
    \end{subequations}
\end{itemize}
Combining all the expressions returns the final form of the integral in Eq.~\eqref{M-integral-1}.

\subsection{Closed-form expressions for the integral functions}
Eq.~\eqref{integral-nth-power} can be further elaborated using Leibniz's rule on the derivative $\dd^n/\dd \mathcal{A}^n$\,, yielding
\begin{equation}
    I_{n}^{(\mathcal{A},\mathcal{B})}(x)
    = (-1)^n \sum_{k=0}^n \frac{n!}{(n-k)!} \sum_{\ell=0}^k \frac{\Gamma(n-\mathcal{B})\, \mathcal{A}^{n-\mathcal{B}-k+\ell-1}}{\ell!(k-\ell)! \, \Gamma(n-\mathcal{B}-k+\ell)} \dv[\ell]{\mathcal{A}}\left[\int_\mathcal{A}^{\mathcal{A}x}\dd z \, e^{-z} z^{\mathcal{B}-n}\right] \ ,\label{integral-nderivative-A}
\end{equation}
where we have defined $z\coloneqq \mathcal{A}\, x$\,.
We can now expand Eq.~\eqref{integral-nderivative-A} singling out the $\ell=0$ and $\ell=1$ contributions
\begin{multline}
    I_n^{(\mathcal{A},\mathcal{B})}(x)=(-1)^n \sum_{k=0}^n \frac{n!\, \Gamma(n-\mathcal{B})}{(n-k)!} \Bigg\{\frac{\mathcal{A}^{n-\mathcal{B}-k-1}}{k! \, \Gamma(n-\mathcal{B}-k)}\left[\gamma(\mathcal{B}-n+1,\mathcal{A}x)- \gamma(\mathcal{B}-n+1,\mathcal{A}) \right] + \\
    + \frac{\mathcal{A}^{-k}}{\Gamma(k)\, \Gamma(n-\mathcal{B}-k+1)}\left[e^{-\mathcal{A}x}x^{\mathcal{B}-n+1}-e^{-\mathcal{A}} \right] 
    + \sum_{\ell=2}^k \frac{ \mathcal{A}^{n-\mathcal{B}-k+\ell-1}}{\ell!(k-\ell)! \, \Gamma(n-\mathcal{B}-k+\ell)} \dv[\ell]{\mathcal{A}}\left[\int_\mathcal{A}^{\mathcal{A}x}\dd z \, e^{-z} z^{\mathcal{B}-n}\right]\Bigg\} \ .
\end{multline}
Now, examining the last term for $\ell \geq 2$\,, we have
\begin{equation}
    \dv[\ell]{\mathcal{A}} \int_\mathcal{A}^{\mathcal{A}x}\dd z \, e^{-z} z^{\mathcal{B}-n} = x^{\ell}\dv[\ell-1]{z}\left( e^{-z} z^{\mathcal{B}-n} \right) -\dv[\ell-1]{\mathcal{A}}\left[ e^{-\mathcal{A}}\mathcal{A}^{\mathcal{B}-n} \right]  \ ,
\end{equation}
where we used that $z$ is linear in $\mathcal{A}$\,. Recalling the definition of the generalized Laguerre polynomial with Rodrigues' formula
\begin{equation}
    L_{m}^{(\alpha)}(z) = \frac{z^{-\alpha} e^{z}}{m!}\dv[m]{z}\left(e^{-z}z^{m+\alpha} \right)  \qq{for} m\in \mathbb{N}_0 \qq{and} \alpha \in \mathbb{R} \ ,
\end{equation}
we can define $m=\ell-1$ and $\alpha=\mathcal{B}-n-\ell+1$ and write
\begin{equation}
    \dv[\ell-1]{\mathcal{A}} \left[ e^{-\mathcal{A}x}(\mathcal{A}x)^{\mathcal{B}-n}x-e^{-\mathcal{A}}\mathcal{A}^{\mathcal{B}-n} \right] 
    = (\ell-1)!\, \mathcal{A} ^{\mathcal{B}-n-\ell+1} \left[  e^{-\mathcal{A}x} x^{\mathcal{B}-n+1} L_{\ell-1}^{(\mathcal{B}-n - \ell +1) }(\mathcal{A}x) - e^{-\mathcal{A}} L_{\ell-1}^{(\mathcal{B}-n - \ell +1) }(\mathcal{A})\right]\ .
\end{equation}
Finally, we have the expression for $I_n^{(\mathcal{A},\mathcal{B})}(x)$ as the finite sum
\begin{align}\label{I-nth}
    I_{n}^{(\mathcal{A},\mathcal{B})}(x)&=(-1)^n \sum_{k=0}^n \frac{n!\, \Gamma(n-\mathcal{B})}{(n-k)!} \Bigg\{\frac{\mathcal{A}^{n-\mathcal{B}-k-1}}{k! \, \Gamma(n-\mathcal{B}-k)}\left[\gamma(\mathcal{B}-n+1,\mathcal{A}x)- \gamma(\mathcal{B}-n+1,\mathcal{A}) \right] + \notag\\
    &\qquad  \qquad + \frac{e^{-\mathcal{A}x} x^{\mathcal{B}-n+1}-e^{-\mathcal{A}} }{(k-1)!\, \Gamma(n-\mathcal{B}-k+1)\mathcal{A}^{k}} + \sum_{\ell=2}^k \frac{e^{-\mathcal{A}x} x^{\mathcal{B}-n+1} L_{\ell-1}^{(\mathcal{B}-n - \ell +1) }(\mathcal{A}x) - e^{-\mathcal{A}} L_{\ell-1}^{(\mathcal{B}-n - \ell +1) }(\mathcal{A})}{\ell\, (k-\ell)! \, \Gamma(n-\mathcal{B}-k+\ell)\mathcal{A}^{k}} \Bigg\}\ .
\end{align}
For completeness, we report the expressions for the integral functions for $n=1, 2$
\begin{subequations}
\begin{align}
    I_1^{(\mathcal{A},\mathcal{B})}(x) &=  \mathcal{A}^{-\mathcal{B}}(1- \mathcal{A} \, \mathcal{B})\left[\gamma(\mathcal{B},\mathcal{A}x)- \gamma(\mathcal{B},\mathcal{A}) \right]   - \frac{e^{-\mathcal{A}x} x^{\mathcal{B}}-e^{\mathcal{A}}}{\mathcal{A}} \ ,\\
    I_2^{(\mathcal{A},\mathcal{B})}(x) &= \mathcal{A}^{-\mathcal{B}-1}\left(\mathcal{A}^2+2\mathcal{A}(1-\mathcal{B})+2 \mathcal{B}(\mathcal{B}-1)\right)\left[ \gamma(\mathcal{B}-1,\mathcal{A}x)-\gamma(\mathcal{B}-1,\mathcal{A})\right] + \notag \\
    &  \qquad \qquad  +\frac{\mathcal{B}(2\mathcal{A}(\mathcal{B}-1) - 1)}{\mathcal{A}^2} \left( e^{-\mathcal{A}x}x^{\mathcal{B}-1} -  e^{-\mathcal{A}}\right) +\frac{e^{-\mathcal{A}} - x^\mathcal{B}\,  e^{-\mathcal{A}x}}{\mathcal{A}}\ .
\end{align}
\end{subequations}

\subsection{Convergence of the sums}\label{convergence}
The solution the mass function \eqref{M-sol-I} is expressed in terms of ${\cal I}(r)$\,, which is given in~\eqref{M-integral-1} and can be re-expressed as a sum of 7 different series.
To ensure the validity of the solution, it is therefore important to check whether the series are pointwise convergent.
The $n$-th summand in ach of these series has a similar structure to the first series, which we computed in Eq.~\eqref{integral-mu2-piece}: it is given by the product of a constant prefactor (determined by properties of the scalar field, e.g.~its mass), times a quadratic combination of the coefficients of the series $S(r)$ in Eq.~\eqref{S definition}, times the $n$-th order integral function. Hence, to test the converge of each such series, we apply the standard root convergence criterion, which requires evaluating
\begin{equation}
    L(x)=\lim_{n \to \infty} \left| c_nI_{n}^{(\mathcal{A},\mathcal{B})}(x)\right|^{1/n} \qq{with} c_n= \sum_{p=0}^n a_p \overline{a}_{n-p} \ .
\end{equation}
The series is guaranteed to converge if $L(x) < 1$\,.
We start by observing that, by the convergence condition, the coefficients $a_n$ are at most rational functions of $n$, so that $ \limsup_{n\to \infty} a_n \sim n^{\alpha}$ with $\alpha \in \mathbb{R}$\,. Hence, the limit of the coefficient $c_n$ is $\limsup_{n\to \infty}|c_n|^{1/n} =1$\,. Meanwhile, the large-$n$ behavior of the integral function~\eqref{I-nth} can be analyzed via Laplace's method~\cite{Bender:1999box}, using the the integral formula for $I_{n}^{(\mathcal{A},\mathcal{B})}$ \eqref{Inth-def} and re-expressing it as 
\begin{equation}
    I_{n}^{(\mathcal{A},\mathcal{B})}(x)= \int_1^x \dd \widetilde{x} \, e^{-\mathcal{A}\widetilde{x} + \mathcal{B}\log(\widetilde{x}) + n \phi(\widetilde{x})} \qq{with} \phi(\widetilde{x}) \coloneqq \log\left(1 -  \frac{1}{\widetilde{x}} \right) \ , 
\end{equation} 
where $\tilde{x}$ is a dummy integration variable. Since $\phi'(\widetilde{x}) > 0$\,, the phase of the integrand is a monotonically increasing function of $\widetilde{x}$\,, and therefore the dominant contributions to the large-$n$ asymptotics of the integral arise from a small neighborhood of its upper integration limit.
Hence, we can Taylor-expand $\phi$ around the point $\widetilde{x} = x$ and define $s = - n \phi^\prime(x)(\widetilde{x}-x)$\,, which gives
\begin{equation}
    I_{n}^{(\mathcal{A},\mathcal{B})}(x) \approx  e^{n \phi(x)}\int_0^{n \phi'(x)x} \frac{\dd s}{n \phi^\prime(x)} \, e^{-\mathcal{A} \left( x+\frac{s}{n \phi^\prime(x)}\right) + \mathcal{B}\log\left( x+\frac{s}{n \phi^\prime(x)}\right) -s} =  \frac{e^{-\mathcal{A}x}x^{\mathcal{B}-1}}{n(x-1)}\left(1-\frac{1}{x} \right)^n \ .
\end{equation}
Taking the $n$-th root of the asymptotic expression and using
$\limsup_{n\to\infty}|c_n|^{1/n}=1$\,, and recalling
$\lim_{n\to\infty}n^{-1/n}=1$\,, yields
\begin{equation}
\limsup_{n\to\infty}
\left|c_n I_n^{(\mathcal A,\mathcal B)}(x)\right|^{1/n}
=
\left(1-\frac{1}{x}\right)
\limsup_{n\to\infty}
\left|
\frac{c_n e^{-\mathcal A x}x^{\mathcal B-1}}
     {n(x-1)}
\right|^{1/n}
=
1-\frac{1}{x}~.
\end{equation}
We conclude that
\begin{equation}
    L(x) = 1-\frac{1}{x}  < 1 ~.
\end{equation}
Hence, the series is absolutely convergent for every finite value of $x$\,. However, the convergence is not uniform on the interval $[1,\infty)$\,, since in the asymptotic limit $x\to\infty$\,, the root test yields $L=1$ and is therefore inconclusive. Hence, to verify the convergence of the integral~\eqref{Itot-def}, we need to rely to a uniform bound on the summability.
But first, let us remark that we can explicitly compute the integral by introducing a shift in the integration variable $\widetilde{x} = 1 + t$\,, obtaining
\begin{equation}
   I_{n}^{(\mathcal{A},\mathcal{B})} = \int_1^\infty \dd \widetilde{x} \, e^{- \mathcal{A} \widetilde{x}} \widetilde{x}^{\mathcal{B}-n} (\widetilde{x}-1)^n = e^{-\mathcal{A}}\int_0^\infty \dd t \, e^{- \mathcal{A}t} (t+1)^{\mathcal{B}-n} t^{n} =  e^{-\mathcal{A}} n! \, U(n+1,\mathcal{B}+2,\mathcal{A}) \ ,
\end{equation}
where $U(a,b,z)$ is the Tricomi function, which is a confluent hypergeometric function of second kind. As for the asymptotic expansion, we have that the stationary point, in the large-$n$ limit, is determined by $t_\ast \sim \sqrt{n}$\,.
Let us now consider the following inequality
\begin{equation}
    \frac{t}{t+1}=1-\frac{1}{t+1} \leq e^{-1/(t+1)} \implies I_{n}^{(\mathcal{A},\mathcal{B})} \leq 
    e^{-\mathcal{A}}\int_0^{\infty} \dd t \, e^{- \mathcal{A}t} (t+1)^{\mathcal{B}} e^{-n/(t+1)} \ . 
\end{equation}
Now, we can split the integration region at $t=\sqrt{n}$ and start considering $t\in[0,\sqrt{n}]$\,, so that $t+1 \leq \sqrt{n}+1 \leq 2 \sqrt{n}$\,, for $n \geq 1$\,, and hence we have $e^{-n/(t+1)} \leq e^{-\sqrt{n}/2}$\,, which yields the following chain of inequalities
\begin{align}\label{bound-lower-saddle}
    e^{-\mathcal{A}}\int_0^{\sqrt{n}} \dd t \, e^{- \mathcal{A}t} (t+1)^{\mathcal{B}-n} t^{n}   
     \leq e^{-\mathcal{A}-\sqrt{n}/2}\int_0^{\sqrt{n}} \dd t \, e^{- \mathcal{A}t} (t+1)^{\mathcal{B}}  
     \leq e^{-\mathcal{A}-\sqrt{n}/2}\int_0^{\sqrt{n}} \dd t \,  (t+1)^{\mathcal{B}}  
     \leq \mathcal{C}_1 n^{(\mathcal{B}+1)/2}e^{-\sqrt{n}/2} \ .
\end{align}
Next, for the upper branch of integration $t > \sqrt{n}$\,, we can simply consider $e^{-n/(t+1)} \leq 1$ and that $(t+1)^\mathcal{B} \leq c_\mathcal{B} \,t^\mathcal{B}$ for $t\geq 1$ and some constant $c_\mathcal{B}$\,. Hence, we obtain the following chain of inequalities
\begin{align}\label{bound-upper-saddle}
    e^{-\mathcal{A}}\int_{\sqrt{n}}^\infty \dd t \, e^{- \mathcal{A}t} (t+1)^{\mathcal{B}-n} t^{n} \leq c_\mathcal{B}\, e^{-\mathcal{A}}\int_{\sqrt{n}}^\infty\dd t \,  e^{- \mathcal{A}t} t^{\mathcal{B}} \leq \mathcal{C}_2  \, e^{-\mathcal{A}\sqrt{n}} \ ,
\end{align}
where in the latter inequality, we used the tail estimate for the incomplete upper Gamma function~\cite{Bender:1999box}. Therefore, we can combine the two bounds, \eqref{bound-lower-saddle} and \eqref{bound-upper-saddle}
\begin{equation}\label{large-n-bound}
    I_{n}^{(\mathcal{A},\mathcal{B})} \leq \mathcal{C}_1 n^{(\mathcal{B}+1)/2}e^{-\sqrt{n}/2} + \mathcal{C}_2  \, e^{-\mathcal{A}\sqrt{n}} \leq \mathcal{C} \, n^{(\mathcal{B}+1)/2} e^{- \mathcal{A} \sqrt{n}} \ .
\end{equation}
Finally, since the majorant on the right hand side of \eqref{large-n-bound} is the general term of a converging series in $n$\,, we  conclude that the expression for $\mathcal{I}_{\rm tot}$\,, given as the $\lim_{r\to \infty} \mathcal{I}(r)$ of the series in Eq.~\eqref{M-integral-1}, is a convergent power series. 

\bibliographystyle{unsrtnat}
\bibliography{bibliography.bib}

\end{document}